\documentclass[aps,prl,twocolumn,nopacs,superscriptaddress]{revtex4}
\usepackage[utf8]{inputenc}
\usepackage[T1]{fontenc}

\usepackage{graphicx}  
\usepackage{dcolumn}   
\usepackage{bm}        
\usepackage{amssymb}   
\usepackage{amsmath}
\usepackage{units}
\usepackage[english]{babel}
\usepackage[dvipsnames]{xcolor}
\usepackage{xfrac}
\usepackage{multirow}
\usepackage{circledsteps}

\usepackage[%
colorlinks=true,
urlcolor=RoyalBlue,
linkcolor=RoyalBlue,
citecolor=RoyalBlue,
]{hyperref}

\usepackage{vmargin}
\setpapersize{A4} \oddsidemargin20mm \textwidth170mm \topmargin10mm
\usepackage{sidecap}
\sidecaptionvpos{figure}{t}  

\usepackage[type1]{libertine}                                        
\usepackage{textcomp}
\usepackage[scaled=.85]{beramono}
\usepackage[libertine,cmintegrals,cmbraces,vvarbb,slantedGreek]{newtxmath}
\usepackage[scr=boondoxo]{mathalfa}
\usepackage{bm}
\usepackage[lf]{carlito}
\usepackage{braket}
\usepackage{bm}

\makeatletter
\DeclareRobustCommand{\cev}[1]{%
  \mathpalette\do@cev{#1}%
}
\newcommand{\do@cev}[2]{%
  \fix@cev{#1}{+}%
  \reflectbox{$\m@th#1\vec{\reflectbox{$\fix@cev{#1}{-}\m@th#1#2\fix@cev{#1}{+}$}}$}%
  \fix@cev{#1}{-}%
}
\newcommand{\fix@cev}[2]{%
  \ifx#1\displaystyle
    \mkern#21mu
  \else
    \ifx#1\textstyle
      \mkern#21mu
    \else
      \ifx#1\scriptstyle
        \mkern#20mu
      \else
        \mkern#20mu
      \fi
    \fi
  \fi
}
\makeatother

\newcommand{\panel}[1]{(#1)}
\newcommand{\panelcaption}[1]{(#1)}
\newcommand{\panelsubcaption}[1]{(#1)}

\begin{document}

\title{Microwave Excitation of Atomic Scale Superconducting Bound States}
\author{Janis Siebrecht}
\affiliation{Max-Planck-Institut f\"ur Festk\"orperforschung, Heisenbergstraße 1, 70569 Stuttgart, Germany}
\author{Haonan Huang}
\affiliation{Max-Planck-Institut f\"ur Festk\"orperforschung, Heisenbergstraße 1, 70569 Stuttgart, Germany}
\author{Piotr Kot}
\affiliation{Max-Planck-Institut f\"ur Festk\"orperforschung, Heisenbergstraße 1, 70569 Stuttgart, Germany}
\author{Robert Drost}
\affiliation{Max-Planck-Institut f\"ur Festk\"orperforschung, Heisenbergstraße 1, 70569 Stuttgart, Germany}
\author{Ciprian Padurariu}
\affiliation{Institut für Komplexe Quantensysteme and IQST, Universität Ulm, Albert-Einstein-Allee 11, 89069 Ulm, Germany}
\author{Bj\"orn Kubala}
\affiliation{Institut für Komplexe Quantensysteme and IQST, Universität Ulm, Albert-Einstein-Allee 11, 89069 Ulm, Germany}
\affiliation{Institute for Quantum Technologies, German Aerospace Center (DLR), Wilhelm-Runge Straße 10, 89081, Ulm, Germany}
\author{Joachim Ankerhold}
\affiliation{Institut für Komplexe Quantensysteme and IQST, Universität Ulm, Albert-Einstein-Allee 11, 89069 Ulm, Germany}
\author{Juan Carlos Cuevas}
\affiliation{Departamento de F\'{\i}sica Te\'orica de la Materia Condensada and Condensed Matter Physics Center (IFIMAC), Universidad Aut\'onoma de Madrid, 28049 Madrid, Spain}
\author{Christian R. Ast}
\email[Corresponding author; electronic address:\ ]{c.ast@fkf.mpg.de}
\affiliation{Max-Planck-Institut f\"ur Festk\"orperforschung, Heisenbergstraße 1, 70569 Stuttgart, Germany}

\date{\today}

\begin{abstract}
Magnetic impurities on superconductors lead to bound states within the superconducting gap, so called Yu-Shiba-Rusinov (YSR) states. They are parity protected, which enhances their lifetime, but makes it more difficult to excite them. Here, we realize the excitation of YSR states by microwaves facilitated by the tunnel coupling to another superconducting electrode in a scanning tunneling microscope (STM). We identify the excitation process through a family of anomalous microwave-assisted tunneling peaks originating from a second-order resonant Andreev process, in which the microwave excites the YSR state triggering a tunneling event transferring a total of two charges.  We vary the amplitude and the frequency of the microwave to identify the energy threshold and the evolution of this excitation process. Our work sets an experimental basis and proof-of-principle for the manipulation of YSR states using microwaves with an outlook towards YSR qubits.
\end{abstract}

\maketitle

Magnetic impurities coupled to a superconductor give rise to Yu-Shiba-Rusinov (YSR) states, which are subgap states protected by parity (even/odd particle number conservation) \cite{YU1965,Shiba1968,rusinov_superconductivity_1969}. They exhibit a variety of interesting phenomena including (but not limited to) their resonant character, which enhances higher order processes in tunneling (Andreev processes) or their parity protection, which enhances their lifetime \cite{Huang2020,Ruby2015,Villas2020}. Comparatively long coherence times can also be expected in YSR states, but work on coherent coupling of YSR states so far has been limited \cite{Huang2020,karan_superconducting_2022}. The first step towards coherent manipulation is the use of microwaves in a tunnel junction, which leads to microwave-assisted tunneling \cite{Kot2020,Peters2020,Royhowdhury2015}. However, parity conservation has to be considered when exciting a YSR state using microwaves. 

Elementary excitations in a superconductor, i.e.\ Bogoliubov quasiparticles, come in pairs due to parity conservation, but only one quasiparticle is needed to excite the YSR state \cite{Balatsky2006}. The second quasiparticle can escape to the continuum, which requires excitation energies of at least the superconducting gap, or through a tunneling contact, where much lower excitation energies are sufficient. A scanning tunneling microscope (STM) provides such a tunneling contact offering the ability to manipulate a YSR state with moderate excitation energies far below the superconducting gap. This makes the STM an ideal platform for the manipulation of YSR states as an extension of nondegenerate Andreev bound states to the atomic scale \cite{janvier_coherent_2015}, which provides a starting point for YSR qubits \cite{tosi_spin-orbit_2019,hays_coherent_2021,matute-cadanas_signatures_2022,pita-vidal_direct_2022}.

Here, we demonstrate the excitation of YSR states using microwaves in the tunnel junction of an STM. We are able to separate different tunneling processes involving the YSR states, which allows us to identify a tunneling process that is only possible through the direct excitation of a YSR state by the microwave. We map out an amplitude threshold that has to be overcome to excite the YSR state. This threshold depends on the applied bias voltage, which allows for great flexibility in different YSR excitation schemes. In this way, we provide a proof of principle for the excitation and manipulation of YSR states by microwaves in the presence of a tunnel junction, which is an important prerequisite for the preparation and control of complex YSR structures, for example, in the context of quantum simulations. 

\begin{figure*}
    \centering
    \includegraphics[width=0.8\textwidth]{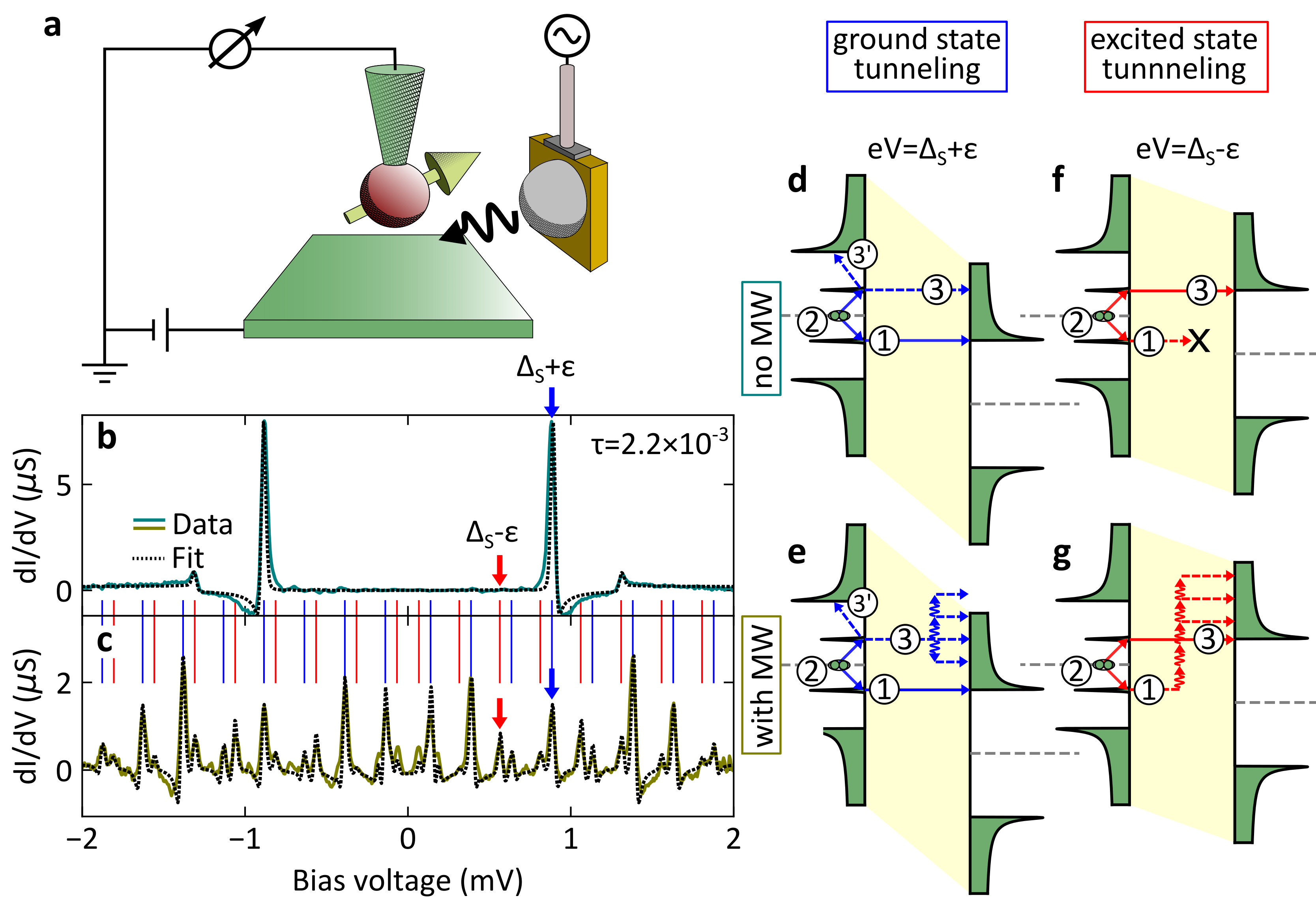}
    \caption{\textbf{Tunneling mechanisms of YSR states under microwave irradiation.} \panelcaption{a} Schematic drawing of the experimental setup. \panelcaption{b} Differential conductance measured without microwaves. Ground state tunneling is indicated by a blue arrow. No excited state tunneling is observed (red arrow). \panelcaption{c} Differential conductance measured with microwaves at 61\,GHz. The energy exchange with the microwave induces replicas. The zero order ground state tunneling is indicated by a blue arrow.  Excited state tunneling induces additional peaks, with the zero order peak indicated by a red arrow. The dashed lines in \panelsubcaption{b} and \panelsubcaption{c} are fits to the data using the full Green's function model and two transport channels (one BCS and one YSR channel (cf.\ \cite{Huang2020,huang_quantum_2020})). \panelcaption{d-g} Schematics illustrating ground state and excited state tunneling processes with and without microwaves. The schematics are drawn for the zero order processes, i.e.\ no net energy quanta transferred. Energy quanta may be absorbed/emitted in steps \Circled{1}/\Circled{3} leading to replicas at different bias voltages.}
    \label{fig:Fig1}
\end{figure*}

We use a scanning tunneling microscope (STM) with an external microwave antenna optimized for operation between 60\,GHz and 90\,GHz \cite{Drost2022}, which is schematically shown in Fig.\ \ref{fig:Fig1}\panel{a}. By controlled dipping of a vanadium tip in a V(100) surface, we create a YSR state at the apex of the tip \cite{Huang2020,huang_quantum_2020}, which is subsequently irradiated by microwaves. In Fig.\ \ref{fig:Fig1}\panel{b}, the differential conductance (green line) through a YSR state in the absence of microwaves is shown. The salient features of the YSR state are two sharp peaks in the superconducting gap at $eV=\pm(\Delta_\text{s}+\varepsilon)$ corresponding to the electron and hole parts of the Bogoliubov quasiparticle ($V$ is the bias voltage, $\Delta_{t,s}$ is the superconducting gap parameter in tip and sample, and $\varepsilon$ is the YSR energy). 

In recent experiments, microwaves have successfully been implemented in STMs with various applications, such as resolving the internal structure of complex tunneling processes. Initial experiments on clean superconductors \cite{Royhowdhury2015} show good agreement with a theory for microwave-assisted tunneling \cite{Tien1963,Falci1991}, which we refer to in the following as Tien-Gordon (TG) theory. This theory predicts the formation of replicas of very sharp spectral features (e.g.\ coherence peaks, YSR states) at integer multiples of $\hbar\omega_\text{r}/e$ weighted by a squared Bessel function ($\omega_\text{r}$ is the microwave radiation frequency, $\hbar$ is Planck's constant, and $e$ is the elementary charge), which depends on the microwave amplitude. Further work has shown that this theory needs to be generalized beyond the tunneling regime for higher order processes such as the Josephson effect or Andreev reflections \cite{Kot2020}. For a non-resonant transfer of $n$ charges, replicas form at multiples of $\hbar\omega_\text{r}/ne$ \cite{Zimmermann1996,cuevas_subharmonic_2002,chauvin_superconducting_2006}. Also, it has been demonstrated that replicas of YSR states can show asymmetries which are not contained within the TG theory \cite{Peters2020}. This was corroborated by a simplified Green's functions approach \cite{Gonzalez2020}. 

The microwaves induce an alternating voltage $V_\text{ac}$ in the tunnel junction, which is on the order of $100$\,$\upmu$V to $10\,$mV. The conductance spectrum with a YSR state irradiated by microwaves at a frequency of $\omega_\text{r}/2\pi = 60.05$\,GHz and an amplitude of $570$\,$\mu$V is shown in Fig.\ \ref{fig:Fig1}\panel{c} (yellow green line). We note that the temperature of the junction only increases by a few mK, which we can safely assume to be constant in line with previous work \cite{Kot2020}. The interaction of the tunneling electrons with the microwave leads to both the absorption and emission of energy quanta by the tunneling electrons in integer multiples of $\hbar\omega_\text{r}=248.3\,\upmu\text{eV}$. In the simplest approximation, this interaction leads to the appearance of replicas of the spectral features in Fig.\ \ref{fig:Fig1}\panel{b}. In Fig.\ \ref{fig:Fig1}\panel{c}, the expected replicas of the YSR states are indicated by blue vertical lines at distances of $248.3\,\upmu\text{V}$. However, we also observe a number of additional peaks marked by the red vertical lines, which appear at $eV=\pm(\Delta_\text{s}-\varepsilon)+n\hbar\omega_\text{r}$, where $n$ is an integer. This might suggest a thermal origin, but the temperature of 560\,mK is very low and no corresponding peak can be seen in the spectrum in the absence of microwaves (cf.\ Fig.\ \ref{fig:Fig1}\panel{b}). 

\begin{figure*}
    \centering
    \includegraphics[width=\textwidth]{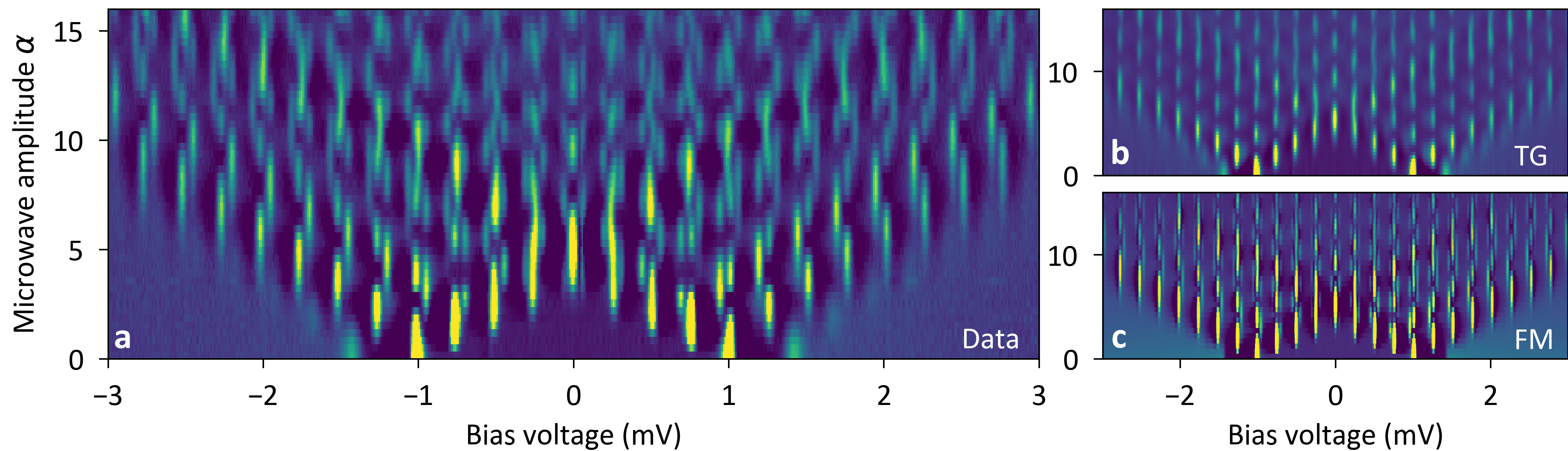}
    \caption{\textbf{Differential Conductance as function of bias of bias voltage and microwave amplitude.} \panelcaption{a} Experimental data measured at a setpoint of 500$\,$pA at 3$\,$mV with a microwave frequency 61$\,$GHz. \panelcaption{b} Calculation based on the spectrum at zero amplitude in panel \panelsubcaption{a} using the Tien-Gordon (TG) model. The features connected to excited state tunneling are missing. \panelcaption{c} Full Green's function model (FM) calculation showing all details as in the experimental data.}
    \label{fig:Fig2}
\end{figure*}

To understand the origin of the different peaks seen in Fig.\ \ref{fig:Fig1}\panel{c}, we present schematics of the underlying tunneling processes in Fig.\ \ref{fig:Fig1}\panel{d}-\panel{g}. To induce tunneling through the YSR state without microwaves, we apply a bias voltage of $eV=\Delta_\text{s}+\varepsilon$ as shown in Fig.\ \ref{fig:Fig1}\panel{d}. To illustrate this, we divide the tunneling process into three steps using the density of states picture. In the first step (labelled \Circled{1}), an electron is transferred across the tunnel junction. In the second step (labelled \Circled{2}) a Cooper pair is split filling the hole, but leaving the YSR state excited. This excited quasiparticle then relaxes into the continuum (step \Circled{3'}) or tunnels across the junction as well (step \Circled{3}). If the tunnel coupling is weak, quasiparticle relaxation in the YSR electrode dominates (step \Circled{3'}). As the tunnel coupling increases, step \Circled{3} becomes dominant transferring a total of two charges across the junction. This step is termed ``resonant Andreev process'' as its tunneling path involves a real state (the YSR state \cite{Yeyati1997,Villas2020,Peters2020}) instead of a virtual state as in conventional Andreev reflections \cite{Ternes2006}. We note that higher-order transfer processes appear in resonant tunneling processes at much lower conductances than for ``conventional'' tunneling, e.g.\ Andreev reflections. Therefore, a theoretical description has to include these processes processes already at a conductance of $2.2\times 10^{-3}G_0$, where Andreev reflections can still be neglected ($G_0=2e^2/h$ is the quantum of conductance). The event illustrated in Fig.\ \ref{fig:Fig1}\panel{d} leads to a spectral peak indicated by the blue arrow in Fig.\ \ref{fig:Fig1}\panel{b}.

In the presence of microwaves, the tunneling process indicated by the blue arrow in the experimental spectrum in Fig.\ \ref{fig:Fig1}\panel{c} is schematically shown in Fig.\ \ref{fig:Fig1}\panel{e}. We first observe the conventional peak to appear at a bias voltage of $eV=\Delta_\text{s}+\varepsilon$, which implies that a total of zero energy quanta are exchanged with the microwave during step \Circled{1}. However, energy quanta can be exchanged during step \Circled{3}, yet without shifting the position of the peak. 

In fact, the peak position only changes if energy quanta are absorbed or emitted during step \Circled{1} such that they appear at different bias voltages $eV=\pm(\Delta_\text{s}+\varepsilon)+n \hbar\omega_\text{r}$ in the spectrum. Other than that, the process is analogous to the tunneling without microwaves (cf.\ Fig.\ \ref{fig:Fig1}\panel{d}). In the following, we consider the two processes involving step \Circled{3} and \Circled{3'} together and refer to this family of peaks as \textit{ground state tunneling}.

The additional peaks seen as red lines in Fig.\ \ref{fig:Fig1}\panel{c} cannot be explained by ground state tunneling (cf.\ Fig.\ \ref{fig:Fig1}\panel{d} and \panel{e}). They can be attributed to processes which originate from tunneling events in absence of microwaves at bias voltages of $eV=\Delta_\text{s}-\varepsilon$ as depicted in Fig.\ \ref{fig:Fig1}\panel{f}, where we would expect them to occur via thermal activation. However, in our experiment, the Boltzmann factor $\exp{(-\frac{\varepsilon}{k_\text{B}T})}$ for a YSR state of energy $\varepsilon=280$\,$\mu$V (for Fig. 2-4) at a temperature of $0.56$\,K predicts a contribution of $0.03$\%, so that thermal excitations are strongly suppressed. Indeed, Fig.\ \ref{fig:Fig1}\panel{b} shows no spectral feature, where the red arrow is pointing. When we turn on the microwaves, a strong and clear peak can be observed at the location of the red arrow, in contrast to a strong peak in presence of microwaves in Fig.\ \ref{fig:Fig1}\panel{c}. In this situation, the microwaves open new transfer channels as delineated in Fig.\ \ref{fig:Fig1}\panel{g}. The absorption of multiple energy quanta during step \Circled{1} induces an excited YSR state (step \Circled{2}) and allows for subsequent relaxation into the continuum through step \Circled{3}. Multiple quanta being absorbed or emitted during process \Circled{3} then lead to a family of additional peaks marked by the red lines in Fig.\ \ref{fig:Fig1}\panel{c} at bias voltages $eV=\pm(\Delta_\text{s}-\varepsilon)+n\hbar\omega_\text{r}$. All the peaks of this family have in common that the excited state is aligned with the coherence peak through the bias voltage modulo an integer number of microwave quanta, which is why call these processes \textit{excited state tunneling}.

\begin{figure}
    \centering
    \includegraphics[width=\columnwidth]{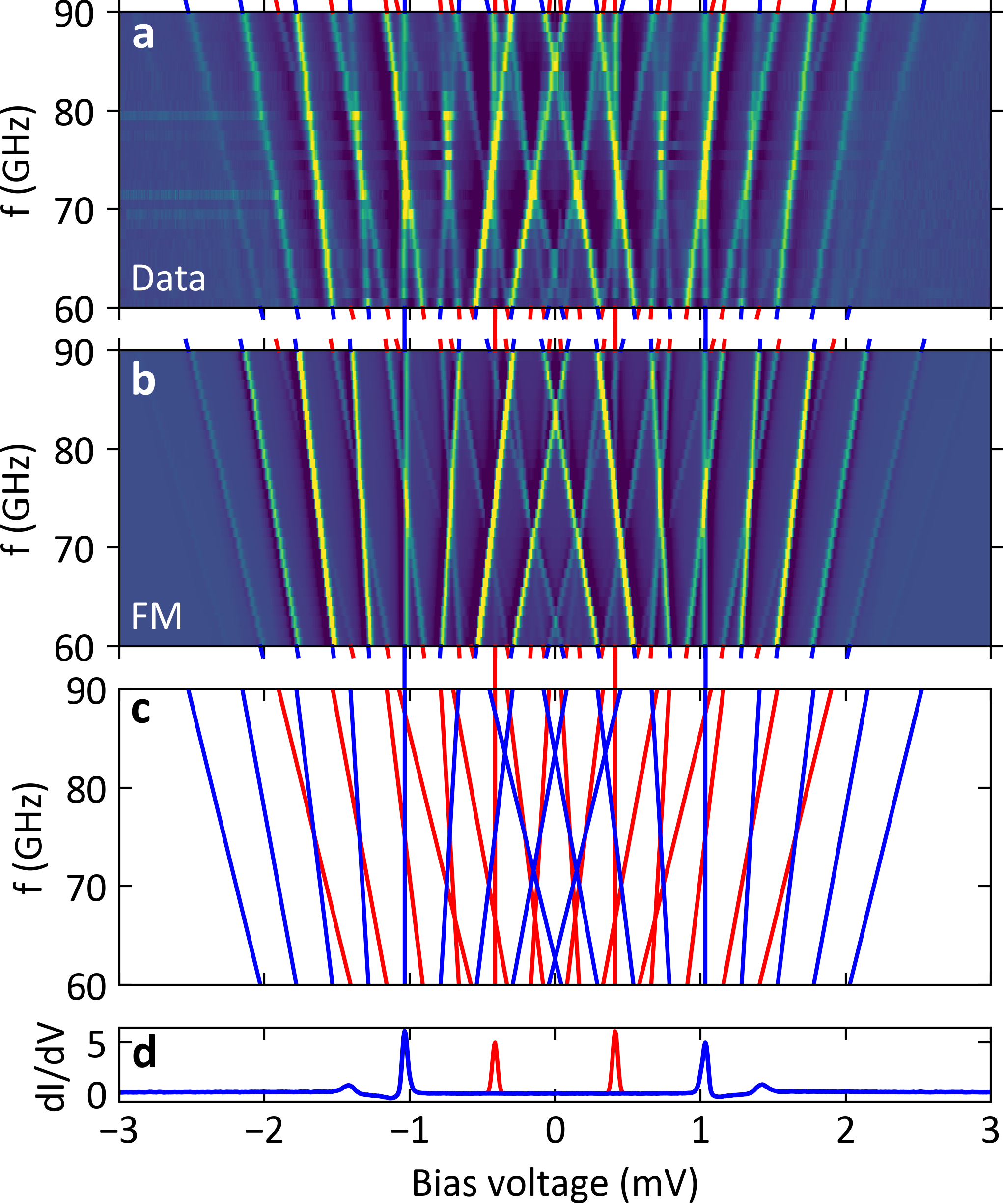}
    \caption{\textbf{Frequency dependence of the spectra at constant microwave amplitude $\alpha$.} \panelcaption{a} Differential conductance spectra measured as function of frequency at constant microwave amplitude $\alpha=\frac{eV_{\text{ac}}}{\hbar \omega_\text{r}}=3$. \panelcaption{b} Calculated spectra in the same range as \panelsubcaption{a} (full model). \panelcaption{c} Theoretical location of normal state replicas (blue) and excited states replicas (red). \panelcaption{d} Base spectrum without microwaves (blue) and excited states added manually (red). The zeroth order replicas (vertical lines) connect the panels.}
    \label{fig:Fig3}
\end{figure}

\begin{figure*}
    \centering
    \includegraphics[width=\textwidth]{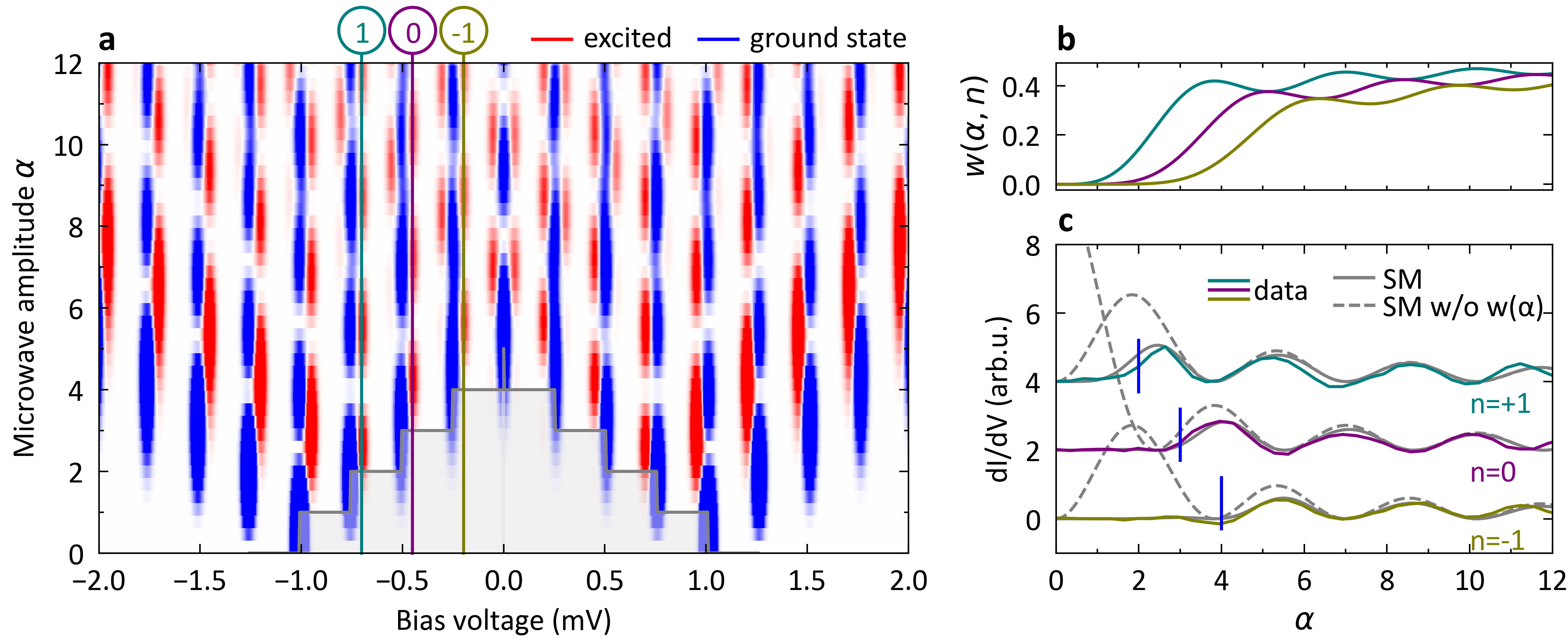}
    \caption{\textbf{Excitation threshold for the YSR state.} \panelcaption{a} Differential conductance calculation (simplified model) to illustrate the origin of the tunneling processes. Ground state tunneling is shown in blue and excited state tunneling in red. \panelcaption{b} The weight function for excited state tunneling for different orders $n=-1,0,1$ as indicated by the vertical lines in \panelsubcaption{a}. The initial threshold is clearly visible. \panelcaption{c} Slices of differential conductance data as function of microwave amplitude for excited state tunneling at different orders $n=-1,0,1$ as indicated by the vertical lines in \panelsubcaption{a}. The simplified model (SM) (solid gray line) fits well to the experimental data and nicely demonstrates the cutoff due to the  threshold at lower amplitudes compared to when the weight function is not considered (gray dashed line). The vertical blue lines indicate the threshold $\alpha = 2,3,4$, respectively.}
    \label{fig:Fig4}
\end{figure*}

In order to understand the evolution of the ground state and excited state tunneling more quantitatively, we measure differential conductance spectra as function of the dimensionless microwave amplitude $\alpha=eV_\text{ac}/\hbar\omega_\text{r}$. Figure \ref{fig:Fig2}\panel{a} shows the differential conductance measured at a microwave frequency of $61$\,GHz and a normal state conductance of $G_\text{N}=2.2\times 10^{-3}G_0$, where $G_0=2e^2/h$ is the quantum of conductance. We can clearly see many well defined peaks, which we will assign to ground state or excited state tunneling in the following. In order to distinguish these peaks, we use the TG model to calculate the expected microwave amplitude dependence from the measured conductance spectrum without microwaves \cite{Tien1963,Royhowdhury2015}
\begin{equation}
    \centering
    I\left(V,\alpha\right)=\sum_n  J_n^2\left(\alpha\right) I^0\!\left(V+ n\hbar\omega_\text{r}/e\right),
    \label{eqn:tg}
\end{equation}
where $J_n(\alpha)$ is the $n$-th order Bessel function of the first kind and $I^0(V)$ is the tunneling current without microwaves. The calculated image starting from the zero amplitude spectrum in Fig.\ \ref{fig:Fig2}\panel{a} is shown in Fig.\ \ref{fig:Fig2}\panel{b}. We note that the TG model does not reproduce all of the experimentally observed peaks. The replicated peaks in Fig.\ \ref{fig:Fig2}\panel{b} are entirely due to ground state tunneling, so that all additional peaks in Fig.\ \ref{fig:Fig2}\panel{a} must be due to excited state tunneling. For comparison, we calculate the data set in Fig.\ \ref{fig:Fig2}\panel{a} using the full Green's function theory taking into account microwaves, higher order tunneling processes (e.g.\ Andreev processes) as well as the interference between them \cite{cuevas_subharmonic_2002} (for details see the Supplementary Information \cite{sinf}). We found that due to the resonant tunneling through the YSR states the interplay between the microwave and the higher order tunneling processes become non-negligible such that approximative calculations fail and the full Green's function model has to be applied (for details see the Supplementary Information \cite{sinf}). The calculation is shown in Fig.\ \ref{fig:Fig2}\panel{c}, which shows excellent agreement with the measured data in Fig.\ \ref{fig:Fig2}\panel{a}. Both ground state and excited state tunneling processes are reproduced with the full Green's function model. 

To substantiate our claim that there are indeed two families of processes, we present frequency dependent differential conductance spectra at a constant dimensionless amplitude of $\alpha=3$ in Fig.\ \ref{fig:Fig3}\panel{a}. The higher the order of the replica, the more tilted the spectral feature will appear in the map. An $n$-th order replica of a feature at $V_0$ moves as $eV=eV_0+n\hbar\omega_\text{r}$. The replica and their dispersion are calculated from the full Green's function theory in Fig.\ \ref{fig:Fig3}\panel{b} as well as presented schematically in  Fig.\ \ref{fig:Fig3}\panel{c}. We can identify four vertical lines corresponding to zero order replicas, marked by the lines connecting panels \panel{a}, \panel{b}, and \panel{c}. The blue and red colors mark ground state tunneling ($eV_0=\pm(\Delta_\text{s}+\varepsilon)$) and excited state tunneling ($eV_0=\pm(\Delta_\text{s}-\varepsilon)$), respectively. We note that at $\alpha=3$, the microwave has enough power to excite the YSR state, such that excited state tunneling becomes possible. If the excited state replica actually did appear in the spectrum without microwaves, which is not the case (cf.\ Fig.\ \ref{fig:Fig1}\panel{b}), the original spectrum would appear as in Fig.\ \ref{fig:Fig3}\panel{d}. In Fig.\ \ref{fig:Fig3}\panel{d}, the excited state tunneling peak is artificially added (red line), where thermal tunneling would appear. However, microwaves could trigger these transfer processes to occur beyond a given threshold as discussed below.

In essence, the breakdown of the simple TG model (Fig.\ \ref{fig:Fig2}\panel{b}) is expected because it leaves the ground state untouched and only considers the spectrum in the absence of microwaves without taking into account processes activated by the microwaves, such as excited state tunneling. The full model (Fig.\ \ref{fig:Fig2}\panel{c}) agrees quantitatively with the experiment (cf. excellent fit in Fig.\ \ref{fig:Fig1}\panel{c}). An intuitive understanding of the mechanism behind excited state tunneling can be derived from a simplified model.  Employing a perturbative approach including second order resonant Andreev processes, the excited state tunneling current $I_\text{ex,e/h}(V,\alpha)$ appears as
\begin{equation}
    \centering
    I_\text{ex,e/h}\left(V,\alpha\right)=\sum_n  w\left(\alpha,n\right) J_n^2\left(\alpha\right) I^0_\text{ex,e/h}\left(V\pm n\hbar\omega_\text{r}/e\right),
    \label{eqn:tgmod}
\end{equation}
where e/h refers to the peak at negative/positive bias voltage $eV=\mp(\Delta_\text{s}-\varepsilon)$. The bare excited state tunneling current $I^0_\text{ex,e/h}(V)$ is replicated by the microwave beyond an amplitude threshold (see below). This is described in the Supplementary Information along with details on the approximations being used \cite{sinf}. We further introduce a weight function
\begin{equation}
    \centering
    w\left(\alpha,n\right)=\sum_{m\geq m_0-n}J^{2}_{m}\left(\alpha\right),
    \label{eqn:wn}
\end{equation}
which sums over all possible energy quanta that can be exchanged during step \Circled{1}, where $m_0=\left \lceil{\frac{2\varepsilon}{\hbar\omega_\text{r}}}\right \rceil $ is the minimum number of quanta needed to excite the YSR state (cf.\ Fig.\ \ref{fig:Fig1}\panel{g}). The excited state tunneling current in Eq.\ \eqref{eqn:tgmod} and the weight function in Eq.\ \eqref{eqn:wn} show that step \Circled{1} in Fig.\ \ref{fig:Fig1}\panel{g} only contributes to the magnitude of the current, but it does not generate any replica. This also explains why the replica are a distance $\hbar\omega_\text{r}/e$ apart despite two charges being transferred in the whole process. A very similar argument can be made for step \Circled{3} of the ground state tunneling in Fig.\ \ref{fig:Fig1}\panel{e}. However, in this case the sum condition in the weight function is $m>n-m_0$, which does not introduce a new threshold, but just leads to a renormalization of the spectral weight. A number of different approximations between the full model and the simplified model in Eq.\ \eqref{eqn:tgmod} can be made, e.g.\ \cite{Gonzalez2020}, which are discussed in the Supplementary Information \cite{sinf}.

In Fig.\ \ref{fig:Fig4}\panel{a}, we separately calculate the ground state and excited state tunneling conductances using Eq.\ \eqref{eqn:tgmod} and the corresponding formula for ground state tunneling \cite{sinf}, which are shown in blue and red, respectively. The stepped shaded area around zero bias voltage represents the threshold amplitude needed to activate excited state tunneling. The weight function for the $n=-1,0,1$ processes marked in Fig.\ \ref{fig:Fig4}\panel{a} are plotted in Fig.\ \ref{fig:Fig4}\panel{b} as function of dimensionless microwave amplitude $\alpha$. We see that the threshold is not a sharp cutoff, but follows the leading edge of the lowest order Bessel function $J_{2,3,4}^2\left(\alpha\right)$, respectively, enabling the process. For $n=-1,0,1$, the threshold is roughly at $\alpha\ge 2,3,4$, respectively, when the weight function becomes significant. To demonstrate the threshold effect of the weight function, we plot the corresponding data from Fig.\ \ref{fig:Fig2}\panel{a} at $n=-1,0,1$ in Fig.\ \ref{fig:Fig4}\panel{c}. The fits are shown with and without the weight function as solid and dashed grey line, respectively. We can directly see how the weight function imposes the threshold for small amplitudes and nicely follows the experimental data. 

At this point, we emphasize that here YSR states are excited using energies much smaller than the minimal energy $\Delta E>\Delta_\text{s}+\varepsilon$, if the YSR state is connected to a tunnel junction. In fact, the excitation energy can be as low as $2\varepsilon$ (cf.\ Fig.\ \ref{fig:Fig1}\panel{g} and Eq.\ \eqref{eqn:tgmod}), which we have demonstrated through the excited state tunneling process and the imposed activation threshold. Even though two electrons are transferred in the resonant tunneling process through the YSR state, the replica are spaced $\hbar\omega_\text{r}/e$ apart instead of $\hbar\omega_\text{r}/2e$ as for conventional Andreev reflections. Hence, the spacing between replica cannot be used for inferring the number of charges being transferred. Our ability to excite YSR states with high precision can now be exploited for direct manipulation protocols. This opens up new possibilities for pump-probe schemes to address the finite lifetime of YSR states.

In summary, we have conducted a proof-of-princple experiment showing that the combination of a tunneling current and microwave radiation can excite YSR states without the need to cross the energy gap. In particular, microwave-assisted tunneling can be used as a tool not only for ground state tunneling, but also for excited state tunneling.  The sub-gap excitation is attractive for future applications (such as information storage) as it does not introduce decoherence by coupling to the continuum to which the YSR state is coupled. Therefore, microwaves could pave the path towards coherent manipulation, similar to ESR-STM \cite{Yang2019} or Andreev qubit architectures \cite{hays_coherent_2021}. Additionally, this work has shown replicas at multiples of $\hbar\omega_\text{r}/e$ as opposed to the $\hbar\omega_\text{r}/2e$ that one would expect for a two-electron process. Pulse schemes or shot noise measurements \cite{Bastiaans2021,Thupakula2022} could shed further light on this process.

\textbf{Acknowledgements}

\noindent The authors thank Klaus Kern for fruitful discussions. This study was funded in part by the ERC Consolidator Grant AbsoluteSpin (Grant No. 681164). JCC thanks the Spanish Ministry of Science and Innovation (Grant No.\ PID2020-114880GB-I00) for financial support as well as the DFG and SFB 1432 for sponsoring his stay at the University of Konstanz as a Mercator Fellow.

\clearpage
\newpage

\onecolumngrid
\begin{center}
\textbf{\large Supplementary Material}
\end{center}
\vspace{1cm}
\twocolumngrid

\setcounter{figure}{0}
\setcounter{table}{0}
\setcounter{equation}{0}
\renewcommand{\thefigure}{S\arabic{figure}}
\renewcommand{\thetable}{S\Roman{table}}
\renewcommand{\theequation}{S\arabic{equation}}

\section{Tip and Sample Preparation}

\begin{figure}
    \centering
    \includegraphics[width=\columnwidth]{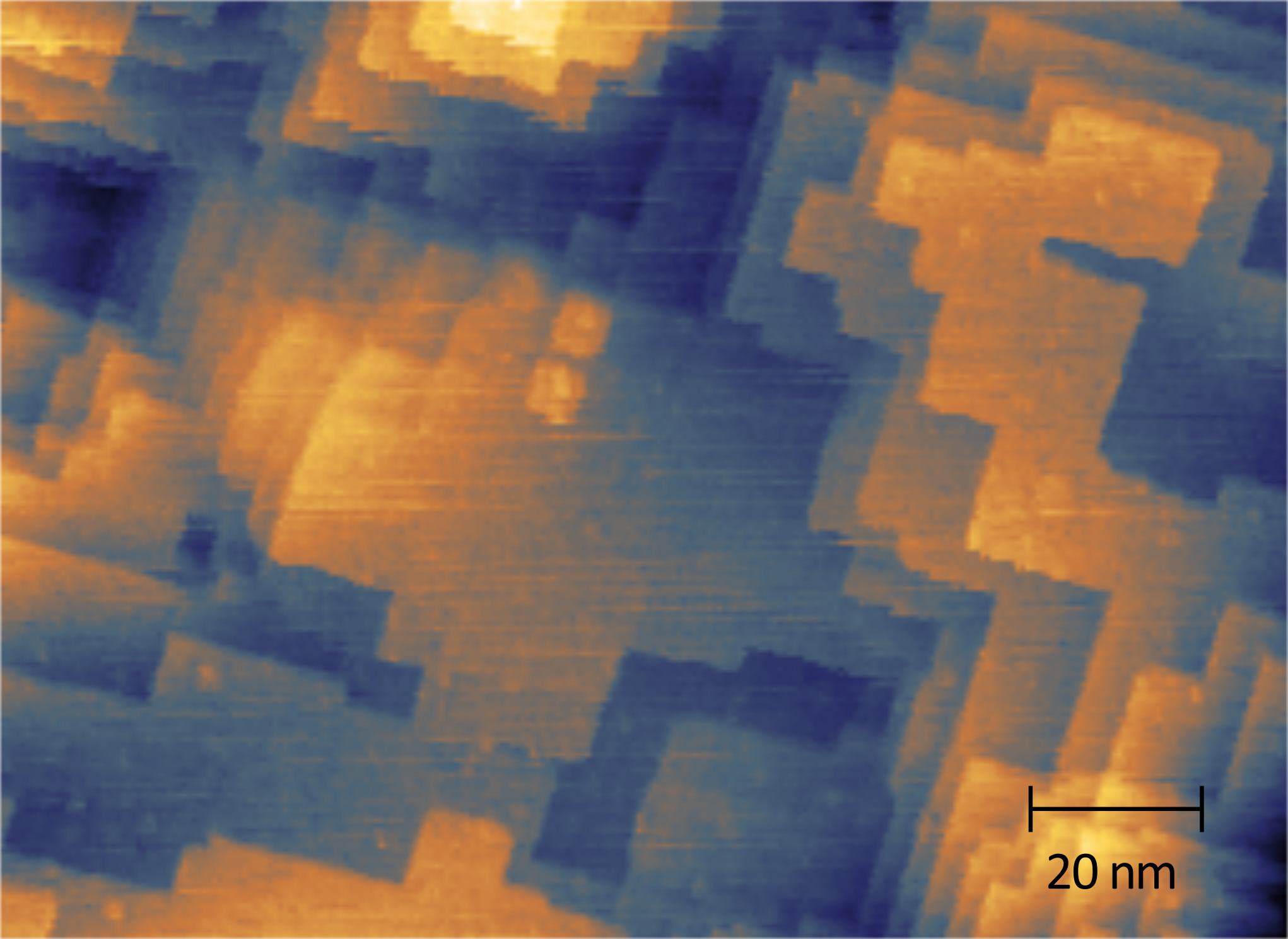}
    \caption{\textbf{Topography of the V(100) surface.} The data was obtained at a set point of 100\,pA with a bias voltage of 3\,mV.}
    \label{fig:vtopo}
\end{figure}

The V(100) sample was cleaned by repeated Argon ion bombardment and annealing to 700$\,^\circ$C. The typical appearance of the surface are square terraces with an oxygen reconstruction as shown in Fig.\ \ref{fig:vtopo} The tip was made superconducting using field emission (40$\,$V bias voltage and 15$\,\mu$A current). By controlled dipping (4$\,$nm dip at $100\,$mV), YSR states were created on the apex of the tip \cite{si_Huang2020,si_karan_superconducting_2022}. For the present system, depending on the exact composition of the apex, the YSR states appear at different energies, allowing us to tune the YSR energy to the relevant frequency range between 60-100$\,$GHz. 

\section{Microwave Transmission}

\begin{figure}
    \centering
    \includegraphics[width=\columnwidth]{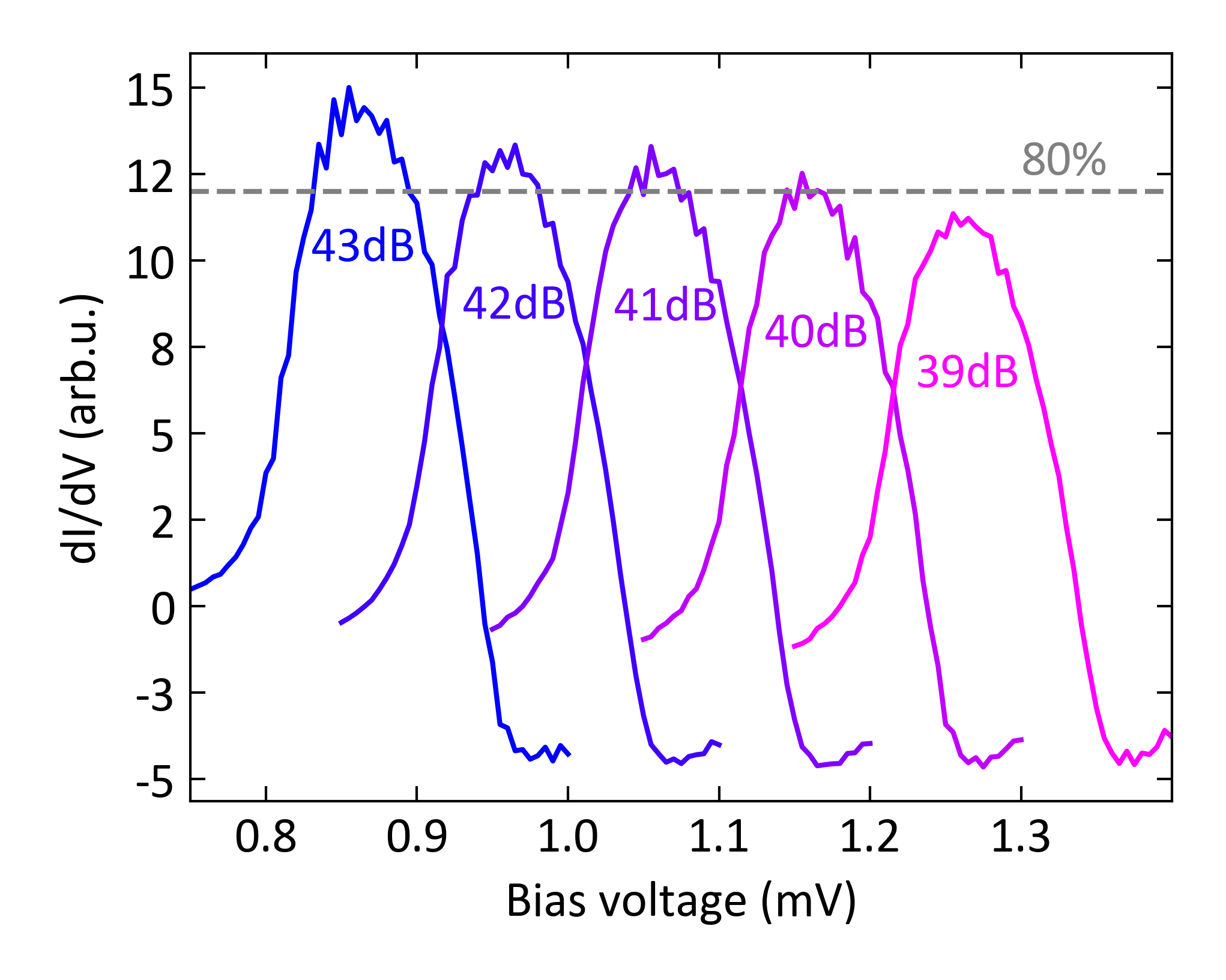}
    \caption{\textbf{Illustration of the algorithm for the transfer function determination.} The plot shows differential conductance spectra of a coherence peak at different values of microwave attenuation. The spectra are offset in voltage for clarity. Starting from a previously determined attenuation, the attenuation is reduced in steps of 1\,dB until the peak height is below the threshold of 80\% of the original peak. The data was measured at a frequency of $61.1$\,GHz, and a setpoint current of 100\,pA at a bias voltage of 3\,mV.}
    \label{fig:tf}
\end{figure}

The microwave setup is similar to the one introduced in Ref.\ \cite{si_drost_combining_2022}. The microwave source is a Keysight 8257D frequency generator (up to 20\,GHz), whose frequency output is multiplied by a factor of six using a Virginia Diodes WR12SGX device. A millimeter wave 511E attenuator is used to tune the attenuation. In the vacuum chamber, we use semirigid Cu coaxial cables, which, starting at the 4\,K stage, is replaced by a superconducting semirigid coaxial cable. Finally, the radiation is transmitted through vacuum to the tunnel junction using a custom made bow-tie antenna on a chip \cite{si_drost_combining_2022}. To measure the transfer function, we use a feedback scheme as shown in Fig.\ \ref{fig:tf}. We broaden the peak by applying a lock-in amplitude and then reduce the attenuation until the peak drops below a threshold of $80$\,$\%$ of its maximum value. Then the ratio of the actual peak height $A_{\omega}$ to the original peak height $A_{0}$ is used to calculated the ac amplitude according to: $\frac{A_{\omega}}{A_{0}}=J_0^2\left(\frac{eV_{\text{ac}}}{\hbar\omega}\right)$, where $J_0$ is the zeroth order Bessel function of the first kind.

\section{Full Green's function theory}

\subsection{General theory}

In this section we show how the general theory of photon-assisted tunneling in superconducting junctions developed in Ref.\ \onlinecite{cuevas_subharmonic_2002} can be adapted to the description of the microwave-assisted tunneling through a magnetic impurity coupled to superconducting leads. Our goal is to calculate the current through a voltage biased superconducting tunnel junction in the presence of a monochromatic radiation of frequency $\omega_\text{r}$. For simplicity, we focus on the case of a single channel contact. We assume that the external radiation produces an effective time-dependent voltage $V(t) = V + V_\textrm{ac} \sin \omega_\text{r} t$. The task now is to extend the theory for multiple Andreev reflections (MARs) to the case of such a time-dependent voltage, for which the so-called Hamiltonian approach is a convenient starting point \cite{si_Cuevas1996}. The irradiated single channel superconducting tunnel junction can be described by means of the following tight-binding-like 
Hamiltonian \cite{si_Cuevas1996}
\begin{equation}
\hat{H} = \hat{H}_\text{L} + \hat{H}_\text{R} + 
\sum_{\sigma} \left\{ t \; c^{\dagger}_{\text{L} \sigma} c_{\text{R} \sigma}
+ t^{*} \; c^{\dagger}_{\text{R} \sigma} c_{\text{L} \sigma} \right\},
\label{eq:ham}
\end{equation}
where $H_\text{L,R}$ are the Hamiltonians following Bardeen-Cooper-Schrieffer (BCS) theory for the isolated electrodes. In the coupling term, L and R stand for 
the outermost sites of each electrode, and $t$ is a hopping parameter describing the coupling between these sites. 
This parameter determines the normal state transmission coefficient $\tau$ of this model in a way that depends on the 
nature of the tunnel junction. For instance, in a tunnel junction formed by two conventional BCS superconductors, the transmission adopts the form 
\begin{equation}
    \tau = \frac{4(t/W)^2}{\left[1+(t/W)^2 \right]^2}\text{, where\ } W=1/\pi\rho_F,
\end{equation}
with $\rho_F$ being the electrodes' density of states at the Fermi level \cite{si_Cuevas1996}.

In this model the current evaluated at the interface between the two electrodes adopts the form
\begin{equation}
I(t) = \frac{i e}{\hbar}  \sum_{\sigma} \left\{ t \langle 
c^{\dagger}_{\text{L} \sigma}(t) c_{\text{R} \sigma}(t) \rangle - 
t^{*} \langle c^{\dagger}_{\text{R} \sigma}(t) c_{\text{L} \sigma}(t) \rangle \right\} .
\label{current}
\end{equation}
The nonequilibrium expectation values in Eq. \eqref{current} can be expressed in terms of the Keldysh-Green 
functions $\hat{G}^{+-}_{ij}$ ($i,j=\text{L,R}$), which in the $2 \times 2$ Nambu representation read
\begin{equation}
\hat{G}^{+-}_{ij}(t,t^{\prime})= i \left( \begin{array}{cc}
\langle c^{\dagger}_{j \uparrow} (t^{\prime}) c_{i \uparrow}(t) \rangle   &
\langle c_{j \downarrow}(t^{\prime}) c_{i \uparrow}(t) \rangle \\
\langle c^{\dagger}_{j \uparrow}(t^{\prime}) c^{\dagger}_{i \downarrow}(t) 
\rangle  & \langle c_{j \downarrow}(t^{\prime}) c^{\dagger}_{i \downarrow}(t)
\rangle \end{array}  \right) .
\end{equation}
Thus, the current can be now written as
\begin{equation}
I(t)  =  \frac{e}{\hbar} \; \mbox{Tr} \left[ \hat{\tau}_3
\left( \hat{t}_\text{LR}(t) \hat{G}^{+-}_\text{RL}(t,t) - \hat{G}^{+-}_\text{LR}(t,t) \hat{t}_\text{RL}(t) \right) \right] ,
\end{equation}
\noindent
where $\hat{\tau}_3$ is the corresponding Pauli matrix in Nambu space, $\mbox{Tr}$ denotes the trace in 
Nambu space and the $\hat{t}$'s are given by
\begin{equation}
\hat{t}_\text{LR}(t) = \hat{t}^{\dagger}_\text{RL}(t) = \left(
\begin{array}{cc}
 t e^{i \phi(t)/2}   &   0      \\
    0                &   -t^{*} e^{-i \phi(t)/2}
\end{array} \right).
\label{eq:v}
\end{equation}
Here, $\phi(t) = \phi_0 + \omega_0 t + 2 \alpha \cos{\omega_\text{r} t}$ is the time-dependent superconducting phase 
difference. In this expression, $\phi_0$ is the dc part of the superconducting phase difference, $\omega_0 = 2eV/ \hbar$
is the Josephson frequency, and the constant $\alpha = eV_\text{ac}/(\hbar \omega_\text{r})$ measures the strength of the 
coupling to the electromagnetic field, and is proportional to the square root of the radiation power. 

Using the relation
\begin{equation}
	e^{i z \cos \phi} = \sum_k i^k J_k(z) e^{i k \phi} ,
\end{equation}
where $J_k(z)$ is the Bessel function of order $k$, one can write the time dependence of the
hopping as follows
\begin{equation}
	t e^{i \phi(t)/2} = t e^{i (\phi_0 + \omega_0 t)/2} \sum_k i^m J_k(\alpha) e^{i k \omega_\text{r} t} .
\end{equation}
In order to determine the Green functions we follow a perturbative scheme and treat the coupling term 
in the Hamiltonian (cf.\ Eq.\ \eqref{eq:ham}) as a perturbation. The unperturbed Green functions $\hat{g}^{r,a}$ correspond to the uncoupled electrodes in equilibrium, where the superscript $r$, $a$ denotes the retarded and advanced components, respectively. Following Ref.~\cite{si_Cuevas1996}, one can express the current in terms of a $T$-matrix, rather than in terms of the Green functions. The $T$-matrix associated to the time-dependent perturbation of Eq.~\eqref{eq:v} is defined as
\begin{equation}
\hat{T}^{r,a} = \hat{t} + \hat{t} \circ \hat{g}^{r,a} \circ \hat{T}^{r,a},
\end{equation}
where the $\circ$ product is a shorthand for integration over intermediate time arguments. With this definition,
it is easy to show that
\begin{eqnarray}
\hat{T}^{r,a}_\text{LR} & = & \hat{t}_\text{LR} + \hat{t}_\text{LR} \circ \hat{g}^{r,a}_\text{R} \circ \hat{t}_\text{RL} \circ \hat{g}^{r,a}_\text{L}  
\circ \hat{T}^{r,a}_\text{LR} , \\ 
\hat{T}^{r,a}_\text{RL} & = & \hat{t}_\text{RL} + \hat{t}_\text{RL} \circ \hat{g}^{r,a}_\text{L} \circ \hat{t}_\text{LR} \circ \hat{g}^{r,a}_\text{R}  
\circ \hat{T}^{r,a}_\text{RL} .
\end{eqnarray}
As shown in Ref.~\cite{si_Cuevas1996}, the current in terms of the T-matrix components reads
\begin{eqnarray}
I(t) = &  \frac{e}{\hbar}  \mbox{Tr} \left[ \hat{\tau}_3
\left( \hat{T}_\text{LR}^r \circ  \hat{g}_\text{R}^{+-} \circ  \hat{T}_\text{RL}^{a} 
\circ \hat{g}_\text{L}^a - \hat{g}_\text{L}^{r} \circ \hat{T}_\text{LR}^r \circ
\hat{g}_\text{R}^{+-} \circ  \hat{T}_\text{RL}^{a} +
\nonumber \right. \right. \\ & \nonumber \left. \left. \hspace{0.5cm}
\hat{g}_\text{R}^{r} \circ \hat{T}_\text{RL}^{r} \circ  \hat{g}_\text{L}^{+-} \circ 
\hat{T}_\text{LR}^{a} - \hat{T}_\text{RL}^{r} \circ  \hat{g}_\text{L}^{+-} \circ
\hat{T}_\text{LR}^{a} \circ \hat{g}_\text{R}^{a} \right) \right]. \\
\end{eqnarray}
In order to solve the $T$-matrix integral equation, it is convenient to Fourier transform with respect to the temporal arguments: 
\begin{equation}
	\hat{T}(t,t^{\prime}) = \frac{1}{2 \pi} \int dE \; 
\int dE^{\prime} \; e^{-i E t/ \hbar} e^{i E^{\prime} t^{\prime}/\hbar} \; \hat{T}(E, E^{\prime}) .
\end{equation}
Due to time dependence of the coupling element (see Eq.~\eqref{eq:v}), one can show that $\hat{T}(E,E^{\prime})$ admits 
the following solution:
\begin{equation} 
    \hat{T}(E,E^{\prime}) = \sum_{n,m} \hat{T}(E, E + neV + m \hbar \omega_\text{r}) 
    \delta(E - E^{\prime} + neV + m \hbar \omega_\text{r}) . 
\end{equation}
Thus, one can finally write down the current as
\begin{equation}
    I(t) = \sum_{n,m} I^m_n \exp \left[ i \left( n \phi_0 + n \omega_0 t + m \omega_\text{r} t \right) \right],
    \label{eq:ftcurrent}
\end{equation}
where the current amplitudes $I^m_n$ can be expressed in terms of the T-matrix Fourier 
components, $\hat{T}^{kl}_{nm}(E) \equiv \hat{T}(E + neV + k 
\hbar \omega_\text{r}, E + meV + l \hbar \omega_\text{r})$, in the following way
\begin{widetext}
\begin{eqnarray}
    I^m_n & = & \frac{e}{h} \int dE \sum_{i,k} \mbox{Tr} \left[ \hat{\tau}_3 \left( \hat{T}^{\substack{r \\ 0k}}_{\text{LR},0i} \hat{g}^{\substack{+- \\ k}}_{\text{R},i}
    \hat{T}^{\substack{a \\ km}}_{\text{RL},in} \hat{g}^{\substack{a \\m}}_{\text{L},n}
    - \hat{g}^{\substack{r \\ 0}}_{\text{L},0} \hat{T}^{\substack{r \\ 0k}}_{\text{LR},0i}
    \hat{g}^{\substack{+- \\ k}}_{\text{R},i} \hat{T}^{\substack{a \\ km}}_{\text{RL},in} +
    \hat{g}^{\substack{r \\ 0}}_{\text{R},0} \hat{T}^{\substack{r \\ 0k}}_{\text{RL},0i}
    \hat{g}^{\substack{+- \\ k}}_{\text{L},i} \hat{T}^{\substack{a \\ km}}_{\text{LR},in}    
    - \hat{T}^{\substack{r \\ 0k}}_{\text{RL},0i} \hat{g}^{\substack{+- \\ k}}_{\text{L},i}
    \hat{T}^{\substack{a \\ km}}_{\text{LR},in} \hat{g}^{\substack{a \\ m}}_{\text{R},n} \right) \right].
    \label{eq:fullgfmodel}
\end{eqnarray}

We are interested in the dc current $I_\textrm{dc}$. In general, this current is the sum of two contributions 
$I_\textrm{dc} = I_\textrm{B} + I_\textrm{Shapiro}$, where $I_\textrm{B} \equiv I^0_0$ (cf.\ Eq.\ \eqref{eq:ftcurrent}) is a background current and $I_\textrm{Shapiro} = \sum_{n,m} I^m_n e^{in\phi_0} \delta(V-V^m_n)$ is the current from Shapiro 
steps contribution at discrete voltages $V^m_n = (m/n) \hbar \omega_\text{r}/2e$. We shall ignore the contribution from the Shapiro steps in the following and focus only on the background current. 

Using the relations
\begin{equation}
	( \hat{T}^{\substack{a \\ kl}}_{\text{RL},ij} )^{\dagger} =  \hat{T}^{\substack{r \\ lk}}_{\text{LR},ji} \;\;\; \mbox{and} 
	\;\;\; \hat{T}^{\substack{r \\ k l}}_{\text{LR},ij} = (-1)^{k-l} \hat{T}^{\substack{a \\ k l}}_{\text{LR},ij} ,
\end{equation}
which can be demonstrated using the corresponding $T$-matrix equations for these components, we can write the dc current exclusively in terms of $\hat{T}^k_i \equiv \hat{T}^{\substack{a \\ k0}}_{\text{LR},i0}$
as follows
\begin{eqnarray} 
I_\textrm{dc} = \frac{2e}{h} \int dE \sum_{i,k} \mbox{Re} \mbox{Tr} \left[ \hat{\tau}_3 
\left( \hat{g}^{\substack{a \\ k}}_{\text{L},i} \hat{T}^{k}_{i}
\hat{g}^{\substack{+-\\0}}_{\text{R},0} \hat{T}^{k \dagger}_{i} -
\hat{T}^{k \dagger}_{i} \hat{g}^{\substack{+-\\k}}_{\text{L},i}
\hat{T}^{k}_{i} \hat{g}^{\substack{a \\0}}_{\text{R},0}
 \right) \right].
 \label{eq-Idc}
\end{eqnarray}
Finally, the $\hat{T}^k_i$ fulfill the following set of linear algebraic equations
\begin{equation}
    \hat{T}^k_i  =  \hat{t}^k_i + 
    \sum_{l} \left\{  \hat{{\cal{E}}}^{kl}_{i,i} \hat{T}^l_i +
    \hat{{\cal{V}}}^{kl}_{i,i+2} \hat{T}^l_{i+2} + \hat{{\cal{V}}}^{kl}_{i,i-2} 
    \hat{T}^l_{i-2} \right\} ,
    \label{T-eq}
\end{equation}
where the different matrix coefficients adopt the following form in terms of the unperturbed Green functions
\begin{eqnarray}
\hat{t}^k_i & = & \frac{t}{2} J_k(\alpha) \left[ i^k (\hat{1} + \hat{\tau}_3
) \; \delta_{i,-1} - (-i)^k (\hat{1} - \hat{\tau}_3) \; \delta_{i,1} \right] 
\nonumber \\
\hat{{\cal{E}}}^{kl}_{i,i} & = & t^2 i^{k+l} \sum_{j} (-1)^j J_{k-j}(\alpha) 
J_{j-l}(\alpha) \left( \begin{array}{cc}
(g^{j}_{R,i+1})_{11} (g^{l}_{L,i})_{11} & 
(g^{j}_{R,i+1})_{11} (g^{l}_{L,i})_{12} \\
(g^{j}_{R,i-1})_{22} (g^{l}_{L,i})_{21} &
(g^{j}_{R,i-1})_{22} (g^{l}_{L,i})_{22}
\end{array} \right) \nonumber \\
\hat{{\cal{V}}}^{kl}_{i,i+2} & = & -t^2 i^{k-l} \sum_{j} J_{k-j}(\alpha) 
J_{j-l}(\alpha) (g^{j}_{R,i+1})_{12} \left( \begin{array}{cc}
(g^{l}_{L,i+2})_{21} &  (g^{l}_{L,i+2})_{22}\\
0  & 0
\end{array} \right) \nonumber  \\
\hat{{\cal{V}}}^{kl}_{i,i-2} & = & -t^2 i^{l-k} \sum_{j} J_{k-j}(\alpha) 
J_{j-l}(\alpha) (g^{j}_{R,i-1})_{21}
\left( \begin{array}{cc}
0 & 0 \\
(g^{l}_{L,i-2})_{11} &  (g^{l}_{L,i-2})_{12}
\end{array} \right) , \nonumber
\end{eqnarray}
\end{widetext}

\noindent
where we have used the shorthand notation $(g^k_{\text{L},i})_{\alpha,\beta} = g^a_{\text{L},{\alpha,\beta}}
(E + ieV + k \hbar \omega_\text{r})$, where $\alpha,\beta = 1,2$ are indexes in Nambu space.

\subsection{Approximations}

In general, one has to solve Eq.\ \eqref{T-eq} numerically to then evaluate the current via Eq.~\eqref{eq-Idc}. However, in low-transmission junctions there are a number of approximations that one can make. In the deep tunnel regime (when the tunnel coupling is the smallest energy scale), one can 
use the following approximation for the solution of Eq.~\eqref{T-eq}:
\begin{equation} \label{approx1}
	\hat{T}^k_i \approx \hat{t}^k_i \;\;\; (i=\pm 1) .
\end{equation}
This leads to the standard Tien-Gordon result for the tunneling of single quasiparticles (see below).

If we want to consider at least the lowest order Andreev reflection, the next approximation is
\begin{eqnarray} \label{approx2}
	\hat{T}^k_1  & \approx & \hat{t}^k_1 + \sum_l \hat{{\cal{V}}}^{kl}_{1,-1} \hat{t}^l_{-1} , \nonumber \\
	\hat{T}^k_{-1} & \approx &  \hat{t}^k_{-1} + \sum_l \hat{{\cal{V}}}^{kl}_{-1,1} \hat{t}^l_{1} .
\end{eqnarray}
Using this approximation in Eq.~\eqref{eq-Idc}, we get the lowest-order approximation for the contributions
of both the quasiparticle current ($|t|^2$) and the Andreev reflection ($|t|^4$). Additionally, one gets terms
like a higher order contribution for the quasiparticle current.

The previous two approximations are perturbative in nature and may lead to divergencies, if they are not properly
regularized. This is what happens, for instance, when there is a bound state inside the gap with a very long lifetime.
In those cases, one can fix that problem by solving the following closed system for $\hat{T}^k_1$ and $\hat{T}^k_{-1}$:
\begin{eqnarray}     
    \hat{T}^k_1 & = & \hat{t}^k_i + \sum_{l} \left\{  \hat{{\cal{E}}}^{kl}_{1,1} \hat{T}^l_1 +
    \hat{{\cal{V}}}^{kl}_{1,-1} \hat{T}^l_{-1} \right\}  \nonumber \\    
    \hat{T}^k_{-1} & = & \hat{t}^k_{-1} + \sum_{l} \left\{  \hat{{\cal{E}}}^{kl}_{-1,-1} \hat{T}^l_1 +
    \hat{{\cal{V}}}^{kl}_{-1,1} \hat{T}^l_{1} \right\} ,
    \label{T-eq-1}
\end{eqnarray}
whose solution is
\begin{widetext}
\begin{eqnarray} \label{T-eq-1-sol}
    \hat{T}_1 & = & \left[1 - \hat{{\cal{E}}}_{1,1} - \hat{{\cal{V}}}_{1,-1} \left[ 1- \hat{{\cal{E}}}_{-1,-1} \right]^{-1}
    \hat{{\cal{V}}}_{-1,1} \right]^{-1} \left( \hat{t}_1 + \hat{{\cal{V}}}_{1,-1} 
    \left[1- \hat{{\cal{E}}}_{-1,-1} \right]^{-1} \hat{t}_{-1} \right) , \nonumber \\
    \hat{T}_{-1} & = & \left[1 - \hat{{\cal{E}}}_{-1,-1} - \hat{{\cal{V}}}_{-1,1} \left[ 1- \hat{{\cal{E}}}_{1,1} \right]^{-1}
    \hat{{\cal{V}}}_{1,-1} \right]^{-1} \left( \hat{t}_{-1} + \hat{{\cal{V}}}_{-1,1} 
    \left[1- \hat{{\cal{E}}}_{1,1} \right]^{-1} \hat{t}_{1} \right) .
\end{eqnarray}
\end{widetext}
Note that in Eq.\ \eqref{T-eq-1-sol} the different matrices have to be understood as big matrices in microwave space. It is worth remarking that this approximation exactly reproduces the results for the YSR problem for the typical transmissions of the experiments.

Actually, there are intermediate approximations that seem to work very well. For instance, to regularize the quasiparticle
term the following approximation suffices
\begin{eqnarray} \label{T-qp}
    \hat{T}_1 & \approx & \left[1 - \hat{{\cal{E}}}_{1,1} \right]^{-1}  \hat{t}_1 , \nonumber \\
    \hat{T}_{-1} & \approx & \left[1 - \hat{{\cal{E}}}_{-1,-1} \right]^{-1} \hat{t}_{-1} ,
\end{eqnarray}
Here, one can ignore the off-diagonal elements (in Nambu space) of $\hat{{\cal{E}}}_{i,i}$. 

The minimal approximation to regularize the Andreev term is given by
\begin{eqnarray} \label{T-AR}
    \hat{T}_1 & \approx & \left[1 - \hat{{\cal{E}}}_{1,1} \right]^{-1} \left( \hat{t}_1 + \hat{{\cal{V}}}_{1,-1} 
    \left[1- \hat{{\cal{E}}}_{-1,-1} \right]^{-1} \hat{t}_{-1} \right),\nonumber \\
    \hat{T}_{-1} & \approx & \left[1 - \hat{{\cal{E}}}_{-1,-1} \right]^{-1} \left( \hat{t}_{-1} + \hat{{\cal{V}}}_{-1,1} 
    \left[1- \hat{{\cal{E}}}_{1,1} \right]^{-1} \hat{t}_{1} \right), 
\end{eqnarray}
where again one can ignore the off-diagonal elements (in Nambu space) of $\hat{{\cal{E}}}_{i,i}$.

\subsection{YSR states + microwaves}

To describe the tunneling through an YSR impurity we use the mean-field Anderson impurity model put forward in 
Refs.~\cite{si_Villas2020,si_Huang2020}. Within this model the Green's functions of the electrodes are given as follows. 
For the left electrode, which is superconducting, we use the standard BCS Green's functions:
\begin{equation}
\label{eq-gBCS}
\hat g_\text{L}(E) = \frac{-\pi N_{0,\text{L}}}{\sqrt{\Delta^2_\text{L} - E^2}} \left[E \tau_0 + \Delta_\text{L} \tau_1 \right] , 
\end{equation}
where $N_{0,\text{L}}$ is the density of states at the Fermi energy of the left electrode in the normal conducting state. On the other hand, the Green functions for the right electrode features a superconducting electrode with the impurity, adopt the form \cite{si_Villas2020}
\begin{widetext}
\begin{equation} \label{GF-imp}
\hat g_\text{R}(E) = \frac{1}{D(E)} \left( \begin{array}{cc}
 E \Gamma_{\rm R} + (E+U-J) \sqrt{\Delta^2_{\rm R} - E^2} & 
 \Gamma_{\rm R} \Delta_{\rm R} \\
 \Gamma_{\rm R} \Delta_{\rm R} &
 E\Gamma_{\rm R} + (E-U-J) \sqrt{\Delta^2_{\rm R} - E^2} \end{array} \right) ,
\end{equation}
where 
\begin{equation}
\label{eq-Dup}
D(E) = 2\Gamma_{\rm R} E (E-J) + \left[(E-J)^2 - U^2 - \Gamma^2_{\rm R} \right] \sqrt{\Delta^2_{\rm R} - E^2} .
\end{equation}
\end{widetext}
Here, we have defined the tunneling rate $\Gamma_{\rm R} = \pi N_{0, \rm R} t^2_{\rm R}$ (a similar rate 
$\Gamma_{\rm L} = \pi N_{0, \rm L} t^2_{\rm L}$ describes the strength of the tip-impurity coupling).

Let us recall that the condition for the appearance of superconducting bound states is $D(E) = 0$. 
In particular, the spin-induced YSR states appear in the limit $|J| \gg \Delta_\mathrm{R}$ (and they are 
inside the gap when also $\Gamma_\mathrm{R} \gg \Delta_\mathrm{R}$). In this case, there is a pair of fully 
spin-polarized YSR bound states at energies $\pm \epsilon$, where
\begin{equation} \label{eq-YSR1}
\epsilon = \Delta_{\rm R} \frac{J^2 - \Gamma^2_{\rm R} - U^2}
{\sqrt{ \left[ \Gamma^2_{\rm R} + (J-U)^2 \right] 
\left[ \Gamma^2_{\rm R} + (J+U)^2 \right]}} ,
\end{equation}
which in the electron-hole symmetric case $U=0$ reduces to 
\begin{equation} \label{eq-YSR2}
\epsilon = \Delta_{\rm R} \frac{J^2 - \Gamma^2_{\rm R}} {J^2 + \Gamma^2_{\rm R}} . 
\end{equation}

In this case, using the approximation of Eq.~\eqref{approx1} we arrive at the following expression for the
quasiparticle current to the lowest order in the tunnel coupling:
\begin{widetext}
\begin{equation} \label{eq-I_qp}
I_\textrm{qp} \approx \frac{4e\pi^2 |t|^2}{h} \sum_k J^2_k(\alpha) \int^{\infty}_{-\infty} \rho_\text{L}(E-eV+k\hbar \omega_\text{r})
\rho_\textrm{R}(E) \left[f(E-eV+\hbar \omega_\text{r}) - f(E) \right] \, dE ,
\end{equation}
\end{widetext}
where $\rho_i(E)$ is the density of states of electrode $i$ and $f(E)$ is the Fermi function. This is simply the 
standard Tien-Gordon result. Thus, because of the presence of YSR states inside the gap (with energy $\epsilon > 0$),
one expects the microwaves to give rise to a series of conductance peaks at $eV = \Delta_\textrm{S} + \epsilon + 
m \hbar \omega_\text{r}$ with a height that should evolve with the microwave power as $J^2_m(\alpha)$.

Using the approximation of Eq.~\eqref{approx2} and selecting the contribution to the resonant Andreev reflection,
we arrive at the following expression for the current due to the resonant Andreev reflection (to lowest order 
in the tunnel coupling):
\begin{widetext}
\begin{eqnarray} \label{eq-I_AR}
I_\textrm{AR} & \approx & \frac{8e\pi^2 |t|^4}{h} \sum_{k,l} J^2_k(\alpha) J^2_l(\alpha) \int^{\infty}_{-\infty} 
\rho_\text{L}(E-eV+k\hbar \omega_\text{r}) \, \rho_\text{L}(E+eV+l\hbar \omega_\text{r}) \, |(g_\text{R})_{12}(E)|^2 \times \nonumber \\
& & \hspace*{4.2cm}\left[f(E-eV+k\hbar \omega_\text{r}) - f(E+eV+l\hbar \omega_\text{r}) \right] \, dE ,
\end{eqnarray}
\end{widetext}
where $(g_\text{R})_{12}(E)$ is the anomalous Green function at the impurity site and it is given in Eq.~\eqref{GF-imp}.

Equations~\eqref{eq-I_qp} and \eqref{eq-I_AR} nicely explain the physics of the experimental observation.
The last remaining thing is to establish what are the simplest expressions that regularize these equations when the YSR states are very long lived. After some careful analysis, we have arrived at the following regularized expressions:
\begin{widetext}
\begin{eqnarray} \label{eq-I_qp-reg}
I^{\rm (reg)}_\textrm{qp} & \approx & \frac{4e\pi^2 |t|^2}{h} \sum_k J^2_k(\alpha) \int^{\infty}_{-\infty} \left\{ 
\frac{\rho_\text{L}(E-eV+k\hbar \omega_\text{r}) \rho_{\textrm{R},1}(E)}{ \big| 1 - |t|^2 J^2_k(\alpha) (g^{\substack{a \\ k}}_{\text{L},-1})_{11} 
(g^{\substack{a \\ 0}}_{\text{R},0})_{11} \big|^2 } \left[f(E-eV+\hbar \omega_\text{r}) - f(E) \right] - \right. \nonumber \\ & & \hspace*{4cm} \left.
\frac{\rho_\text{L}(E+eV+k\hbar \omega_\text{r}) \rho_{\textrm{R},2}(E)}{ \big| 1 - |t|^2 J^2_k(\alpha) (g^{\substack{a \\ k}}_{\text{L},1})_{22} 
(g^{\substack{a \\ 0}}_{\text{R},0})_{22} \big|^2 } \left[f(E+eV+\hbar \omega_\text{r}) - f(E) \right] \right\} \, dE ,
\end{eqnarray}
where $\rho_{\textrm{R},i}(E) = (1/\pi) \mbox{Im} \left\{ (g^{\substack{a \\ 0}}_{\text{R},0})_{ii} \right\}$ $(i=1,2)$, 
$[\rho_{\textrm{R}}(E) = \rho_{\textrm{R},1}(E) + \rho_{\textrm{R},2}(-E)]$, and

\begin{eqnarray} \label{eq-I_AR-reg}
    I^{\rm (reg)}_\textrm{AR} & \approx & \frac{8e\pi^2 |t|^4}{h} \sum_{k,l} J^2_k(\alpha) J^2_l(\alpha) \int^{\infty}_{-\infty} 
    \frac{\rho_\text{L}(E-eV+k\hbar \omega_\text{r}) \, \rho_\text{L}(E+eV+l\hbar \omega_\text{r}) \, |(g_\text{R})_{12}(E)|^2}
    {\big| \left[ 1 - |t|^2 J^2_k(\alpha) (g^{\substack{a \\ k}}_{\text{L},-1})_{11} (g^{\substack{a \\ 0}}_{\text{R},0})_{11} \right] 
    \left[ 1 - |t|^2 J^2_l(\alpha) (g^{\substack{a \\ l}}_{\text{L},1})_{22} 
    (g^{\substack{a \\ 0}}_{\text{R},0})_{22} \right] \big|^2 } \times \nonumber \\
    & & \hspace*{4.7cm}\left[f(E-eV+k\hbar \omega_\text{r}) - f(E+eV+l\hbar \omega_\text{r}) \right] \, dE.
    \label{eq:arreg}
\end{eqnarray}
\end{widetext}
Notice that the only difference with respect to the perturbative results above is the presence of a denominator that regularizes the eventual divergencies. For a comparison of these approximations with the full Green's function model, see Section \ref{sec:valmod}.

\begin{figure}
    \includegraphics[width=\columnwidth]{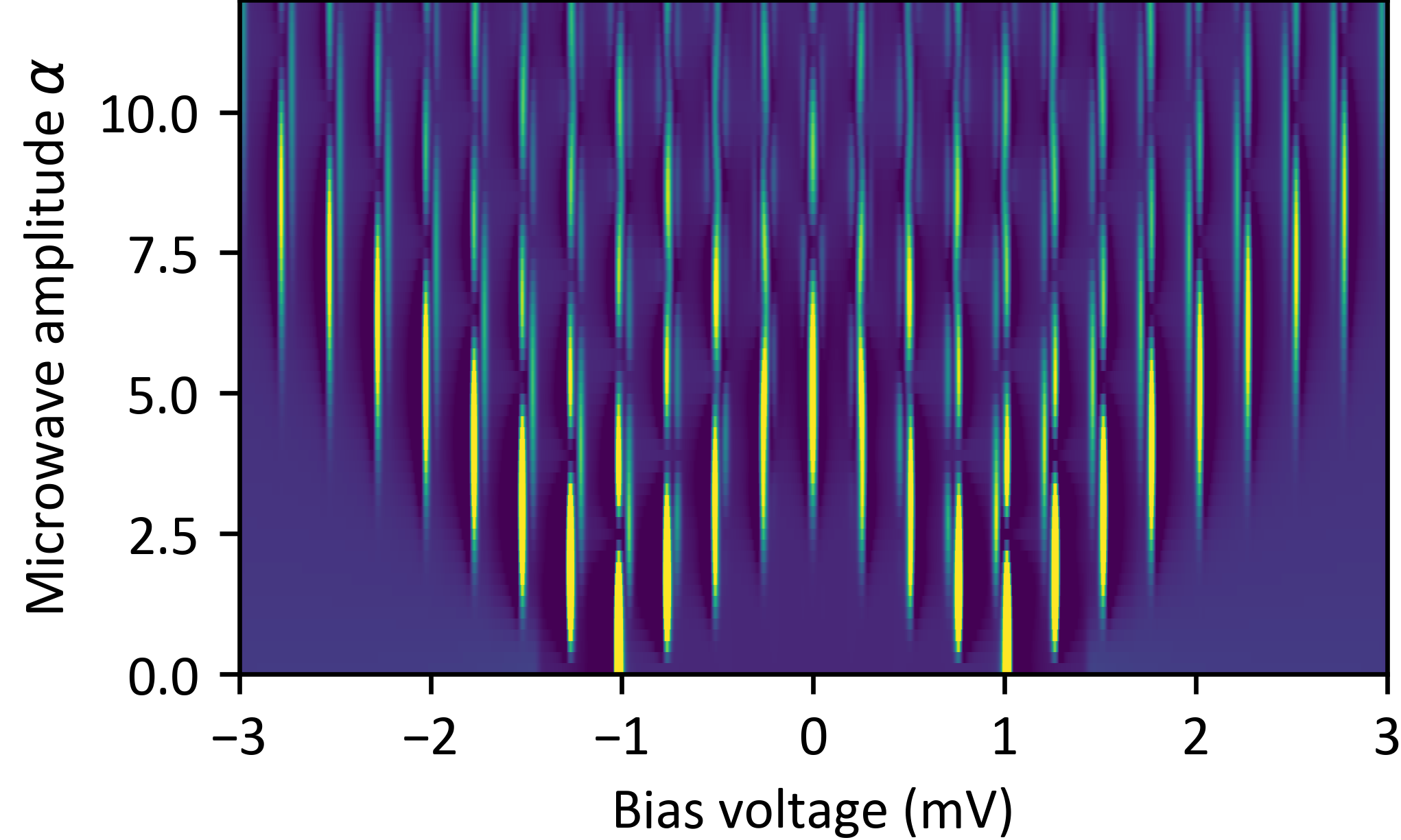}
    \centering
    \caption{\textbf{Calculated differential conductance map as function of microwave amplitude} Calculation of the YSR state replica as function of bias voltage and microwave amplitude using the regularized quasiparticle and Andreev currents (Eq.\ \eqref{eq-I_qp-reg} and \eqref{eq-I_AR-reg}). The excited state tunneling is clearly visible for higher amplitudes.}
    \label{fig-example}
\end{figure}

In Fig.~\ref{fig-example}, we show an example of the differential conductance as a function of the bias voltage $V$ and $\alpha$ computed with Eqs.~\eqref{eq-I_qp-reg} and \eqref{eq-I_AR-reg}, i.e. we computed the total current as the sum of those two contributions. The parameters are given in Table \ref{tab:params}. As one can see, these equations reproduce all the salient features of the experiment.

Finally, to make contact with recent results \cite{si_Gonzalez2020}, we define the energy-dependent
tunneling rates for electrons $\Gamma_\textrm{e}$ and holes $\Gamma_\textrm{h}$ as
\begin{eqnarray}
\Gamma_\textrm{e}(E) & = & 2 \pi |t|^2 \rho_\text{L}(E) \tilde u^2 , \\
\Gamma_\textrm{h}(E) & = & 2 \pi |t|^2 \rho_\text{L}(E) \tilde v^2 ,	
\end{eqnarray}
where the coherent factors $\tilde u^2$ and $\tilde v^2$ in our model are given by
\begin{eqnarray}
\tilde u^2 & = & \frac{2 \Delta_\textrm{R} J \Gamma^2_\textrm{R}}{\left[ \Gamma^2_\textrm{R} + (J+U)^2 \right]}
\frac{1}{\sqrt{\left[ \Gamma^2_\textrm{R} + (J+U)^2 \right] \left[ \Gamma^2_\textrm{R} + (J-U)^2 \right]}} , \nonumber \\
\tilde v^2 & = & \frac{2 \Delta_\textrm{R} J \Gamma^2_\textrm{R}}{\left[ \Gamma^2_\textrm{R} + (J-U)^2 \right]}
\frac{1}{\sqrt{\left[ \Gamma^2_\textrm{R} + (J+U)^2 \right] \left[ \Gamma^2_\textrm{R} + (J-U)^2 \right]}} . \nonumber
\end{eqnarray}
With these definitions, the anomalous Green's function in the impurity can be approximated by (for energies close
to the YSR energy)
\begin{equation}
|(g_\text{R})_{12}(E)|^2 \approx \frac{\tilde u^2 \tilde v^2}{(E- \epsilon)^2 + \eta^2_\textrm{R}} .	
\end{equation}
Thus, the perturbative expression for the current contribution of the resonant AR of Eq.~\eqref{eq-I_AR} becomes

\begin{widetext}
\begin{equation} 
I_\textrm{AR} \approx \frac{2e}{h} \sum_{k,l} J^2_k(\alpha) J^2_l(\alpha) \times 
\int^{\infty}_{-\infty} \frac{\Gamma_\textrm{e}(E-eV-k\hbar \omega_\text{r}) \, \Gamma_\textrm{h}(E+eV+l\hbar \omega_\text{r})}
{(E- \epsilon)^2 + \eta^2_\textrm{R}} \left[f(E-eV-k\hbar \omega_\text{r}) - f(E+eV+l\hbar \omega_\text{r}) \right] \, dE .
\label{eqn:fullexpress}
\end{equation}
\end{widetext}
This expression has to be compared with Eq.~(40) in Ref.~\cite{si_Gonzalez2020}. A list of parameters that were used to obtain the calculated spectra in the different figures is given in Table \ref{tab:params}.

\section{Validity of the models}
\label{sec:valmod}

Resonant tunneling easily involves higher order tunneling long before higher orders become significant in nonresonant tunneling. This is particularly the case when resonant tunneling is combined with the interaction with microwaves. We, therefore, evaluate up to which transparencies such approximations are valid in different models. Here, we compare three models:
\begin{enumerate}
\item Full Green's functions model (Eq.\ \eqref{eq-Idc} with Eq.\ \eqref{T-eq}, black lines)
\item Green's function model to first order in Andreev reflections (Eq.\ \eqref{eq-Idc} with Eq.\ \eqref{T-eq-1-sol}, blue lines)
\item Regularized Andreev model (Eq.\ \eqref{eq-I_qp-reg} and \eqref{eq-I_AR-reg}, red lines)
\end{enumerate}
To assess the agreement of model A referenced to model B with functions $f_\text{A}(x)$ and $f_\text{B}(x)$, we evaluate the mean squared deviation referenced to the mean squared deviation from zero:
\begin{equation}
\chi^2=\frac{\int \left(f_\text{A}(x)-f_\text{B}(x)\right)^2 \text{d}x}{\int f_\text{B}(x)^2 \text{d}dx}
\end{equation}

\begin{figure}
    \centering
    \includegraphics[width=\columnwidth]{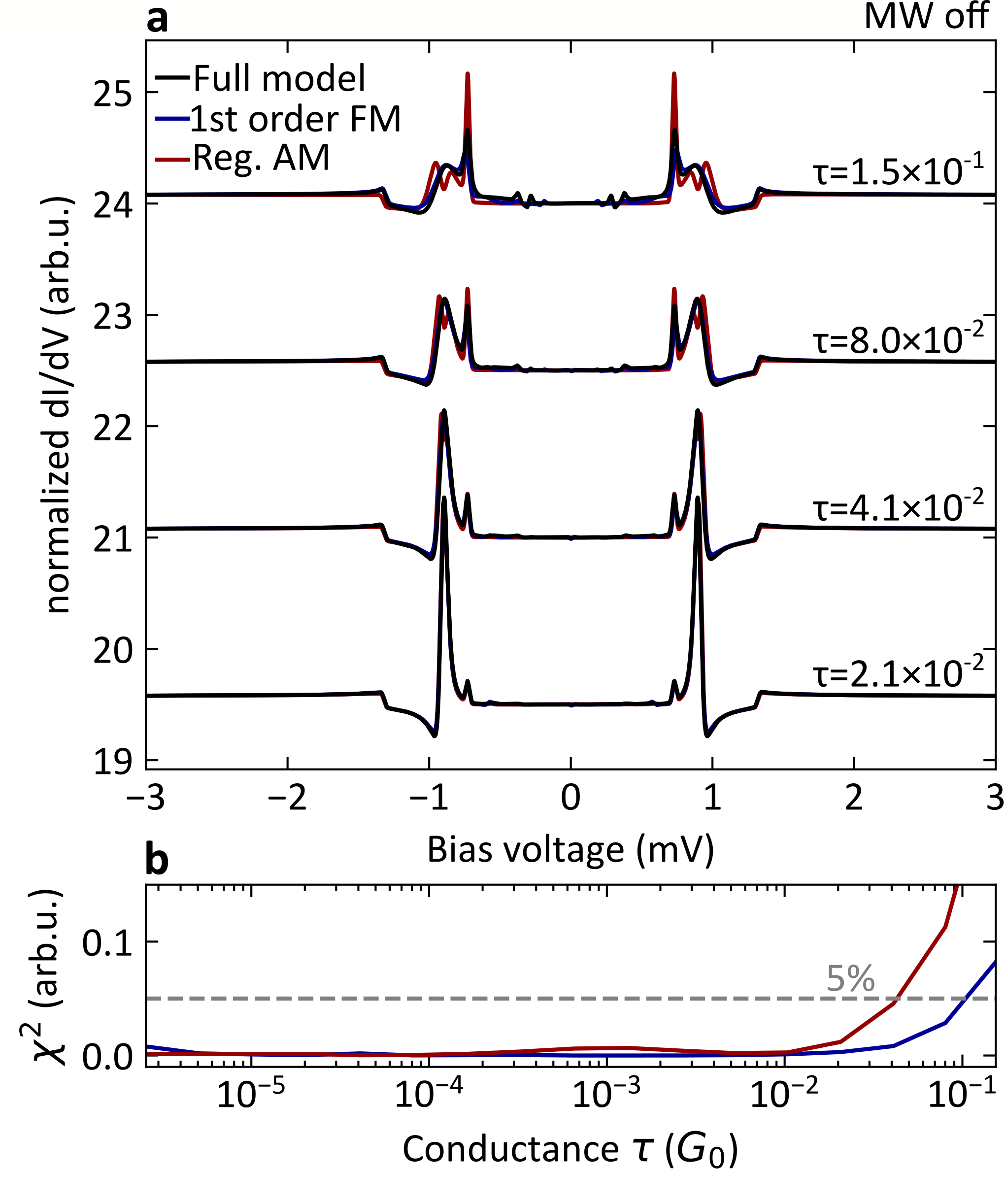}
    \caption{\textbf{Comparison of the different models without microwaves.} \panelcaption{a} Differential conductance spectra calculated using the different models as function of selected junction transparencies $\tau$ with the microwaves turned off. The full Green's function model is labeled ``Full model'', the Green's function model is labeled ``1st order FM'', and the regularized Andreev model is labeled ``Reg.\ AM''. The regularized Andreev model is calculated from the sum of the quasiparticle current and the Andreev current. \panelcaption{b} Deviations of the approximations from the full Green's function model as function of junction transparency $\tau$. Both approximations only fail for high transparencies when higher order processes become relevant also for nonresonant tunneling processes.}
    \label{fig:evolutionnomw}
\end{figure}

We plot the evolution of the spectrum without microwaves as a function of conductance for each model in Fig.\ \ref{fig:evolutionnomw}\panel{a}. Every spectrum is normalized by the normal state tunneling conductance. For each spectrum, we calculated the deviation referenced to the full Green's function model and plot this in Fig.\ \ref{fig:evolutionnomw}\panel{b}.  The regularized Andreev model deviates by more than 5\% from the full calculation at a transparency of $4\times 10^{-2}$, whereas the deviations of the first order model only become relevant at a transparency of $1\times 10^{-1}$. 

\begin{figure}
    \centering
    \includegraphics[width=\columnwidth]{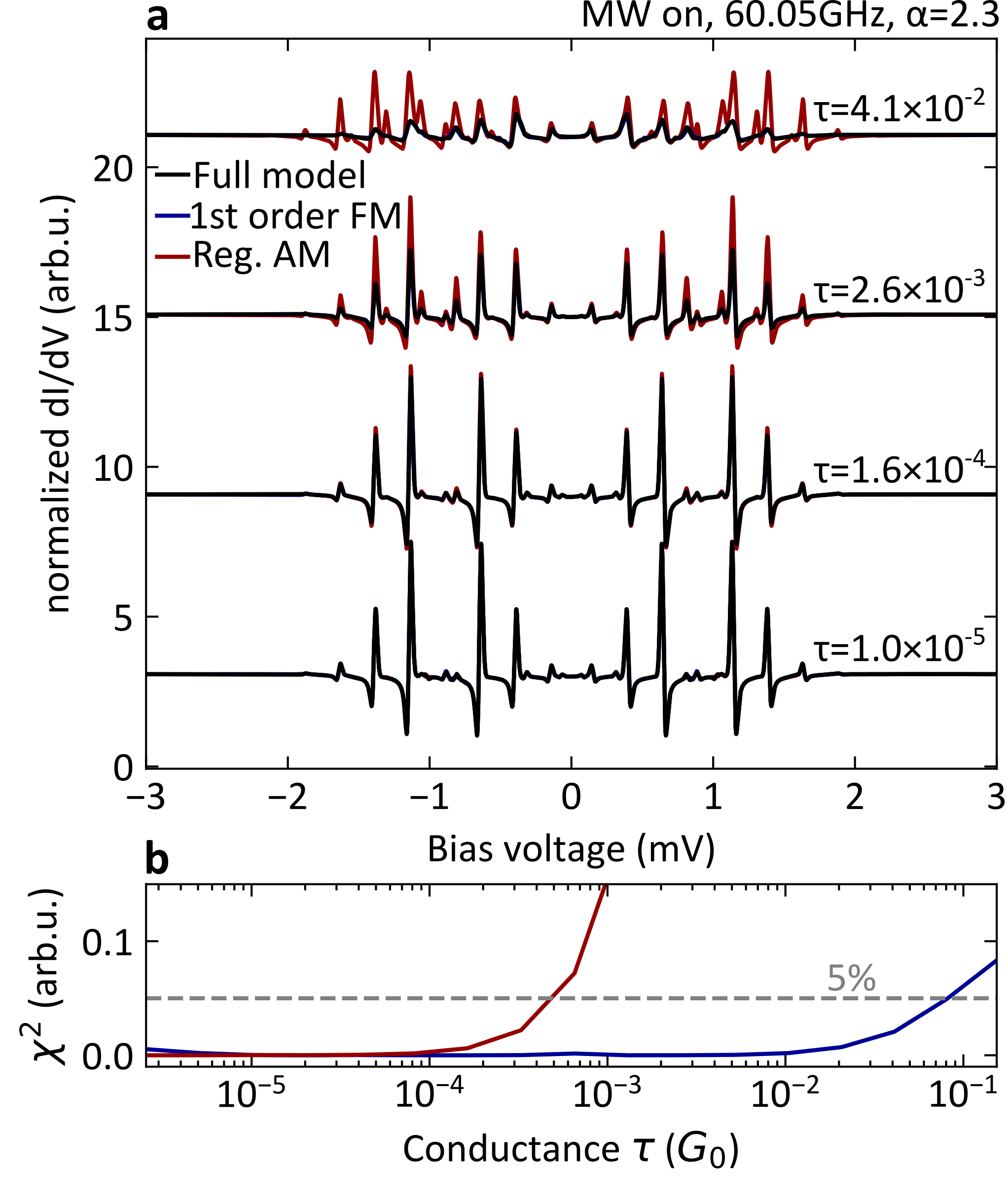}
    \caption{\textbf{Comparison of the different models with microwaves.} \panelcaption{a} Differential conductance spectra calculated using the different models as function of selected junction transparencies $\tau$ with the microwaves turned on. The full Green's function model is labeled ``Full model'', the Green's function model is labeled ``1st order FM'', and the regularized Andreev model is labeled ``Reg.\ AM''. The regularized Andreev model is calculated from the sum of the quasiparticle current and the Andreev current. \panelcaption{b} Deviations of the approximations from the full Green's function model as function of junction transparency $\tau$. The first order model fails at somewhat lower transparencies than without microwaves, but the regularized model fails for two orders of magnitude lower transparencies than before. This indicates that interactions/interference between the resonant processes and the microwaves cannot be neglected for a quantitative and even a qualitative agreement. } 
    \label{fig:evolutionwithmw}
\end{figure}

In contrast to that, when the microwaves are included, the three models become inconsistent much faster. Figure \ref{fig:evolutionwithmw}\panel{a} shows four spectra for each model calculated at different conductances. At the highest conductance, the regularized Andreev model shows deviations in peak height, amplitude and even peak position. The first order approximation still performs much better. This means that the interference of higher order processes is crucial to properly describe the spectrum under microwave irradiation. In Fig.\ \ref{fig:evolutionwithmw}\panel{b}, the deviations start about two orders of magnitude sooner for the regularized Andreev model, which crosses the 5\% mark at a transparency of $5\times 10^{-4}$. The first order approximation crosses the 5\% mark at a transparency of $8\times 10^{-2}$.


The comparison of the behaviour with and without microwaves leads to an important conclusion for the experimental data. Firstly, for measurements without microwaves, the three models remain consistent up to about $\tau=2\times 10^{-2}$, which corresponds to roughly 12\,nA for a set point bias voltage of 4\,mV. This means that for the lifetime broadening of 0.6\,$\upmu$eV (cf.\ Table \ref{tab:params}) used here and for typical setpoint currents of $O(100$\,pA$)$, higher order contributions are not relevant. If the lifetime broadening is smaller, higher order contributions will become relevant at lower transparencies (i.e.\ smaller setpoint currents) \cite{si_Ruby2015,si_Huang2020}. In contrast to that, the regularized Andreev model with microwaves shows significant disagreement already at a transparency of $5\times 10^{-4}$, corresponding to about $150$\,pA. This means that for typical measurements with microwaves and YSR states a full Green's function approach is necessary. The resonances due to the interactions with the microwave lead to a failure of the lowest order approximation \cite{si_Gonzalez2020}. A list of parameters that were used to obtain the calculated spectra in the different figures (both main text and supplementary information) is given in Table \ref{tab:params}.

\begin{table}
    \begin{center}
    \begin{tabular}{l|lllllrlll}
         Figure & $\Delta_\text{L}$&$\Delta_\text{R}$&$\eta_\text{L}$&$\eta_\text{R}$&$J$&$U$&$\Gamma_\text{L}$&$\hbar\omega_\text{r}$&$\alpha$  \\
        \hline\hline
        1b &$0.73$&$0.59$&$0.1$&$0.1$&$75.5$&$0$&$0.07$&0&$0$\\
        1c &$0.73$&$0.59$&$0.1$&$0.1$&$75.5$&$0$&$0.07$&$0.248$&$3.5$\\
        2c &$0.74$&$0.69$&$0.1$&$0.1$&$68.0$&$25$&$0.04$&$0.252$&$\sim$\\
        3b &$0.73$&$0.69$&$0.1$&$0.1$&$64.5$&$25$&$0.04$&$\sim$&$3$\\
        S3 &$0.73$&$0.69$&$0.1$&$0.1$&$64.5$&$25$&$0.04$&$0.252$&$\sim$\\
        S4 &$0.73$&$0.59$&$0.1$&$0.6$&$75.3$&$0$&$\sim$&0&0\\
        S5 &$0.73$&$0.59$&$0.1$&$0.6$&$75.3$&$0$&$\sim$&$0.248$&2.3
    \end{tabular}
    \end{center}
    \caption{\textbf{Fitting Parameters} Table of fit parameters that were used to calculate the spectra in the corresponding figures. The parameters are given in meV, except for $\alpha$, which is dimensionless, and $\eta_\text{L,R}$, which is measured in $\upmu$eV. The temperature was set to be $0.56$\,K, the coupling of the impurity to the substrate $\Gamma_\text{R}=100$\,meV, and an overall Gaussian broadening was chosen to be $12.5\,\upmu$eV. The right electrode (R) carries the YSR state, while the left electrode (L) features an empty gap. For the fit in Fig.\ 1(b), the channel transmissions are $\tau_{\textrm{YSR}}=140\,\text{nS}=1.8\times 10^{-3}\,G_0$ and $\tau_{\textrm{BCS}}=30\,\text{nS}=3.9\times 10^{-4}\,G_0$ for the BCS and the YSR channel, respectively.}
    \label{tab:params}
\end{table}

\section{Derivation of the simplified ground state and excited state tunneling model}

\label{sec:resar}
In the following, we will derive a simplified model to highlight the roles of the replicas in step \Circled{1} and step \Circled{3} of ground state and excited state tunneling (cf.\ schematic in Fig.\ 1\panel{e} and \panel{g} of the main text). We simplify the tunneling and focus on the interplay of the Bessel functions and exchange of energy quanta. We start with Eq.\ \eqref{eqn:fullexpress}. In the case of long-lived YSR states, i.e.\ very small $\eta_S$, we can approximate the Lorentzian of the YSR state by a Dirac delta-function $\frac{1}{(E-\epsilon)^2+\eta_\text{R}^2}\approx\frac{\pi}{\eta_\text{R}}\delta(E-\epsilon)$. This step solves the integral in Eq.\ \eqref{eqn:fullexpress} and the Andreev current becomes
\begin{align}    
    I_\textrm{AR}\approx& \frac{e}{\hbar\eta_\text{R}} \sum_{k,l} J^2_k(\alpha) J^2_l(\alpha) \Gamma_\textrm{e}(\epsilon-eV-k\hbar \omega_\text{r}) \, \Gamma_\textrm{h}(\epsilon+eV+l\hbar \omega_\text{r})\nonumber \\ 
    & \times \left[f(\epsilon-eV-k\hbar \omega_\text{r}) - f(\epsilon+eV+l\hbar \omega_\text{r}) \right]
    \label{eq:ardelta}
\end{align}
Each tunneling rate $\Gamma_\text{e,h}$ has two peaks at $\pm \Delta$, such that we have a total of four peaks in the spectrum. To separate out these peaks, we use the Heaviside step function $\theta (E)$ to define $\Gamma_\text{e,h}^{\pm}(E)=\theta(\pm E)\Gamma_\text{e,h}(E)$ so that we can split $\Gamma_\text{e,h}(E)$ into
\begin{equation}
    \Gamma_\text{e,h}(E)=\Gamma_\text{e,h}^+(E)+\Gamma_\text{e,h}^-(E).
\end{equation}
Without microwaves, the four principal peaks correspond to two ground state and two excited state tunneling peaks at $eV=\pm(\Delta + \epsilon)$ and $eV=\pm(\Delta - \epsilon)$, respectively. The derivations for all of these four peaks are very similar, so that in the following we derive the behavior for one peak, which can be easily extended to the other peaks.

\subsection{Derivation for excited state tunneling electron peak}

These peaks are located at bias voltages of $eV=-(\Delta-\epsilon)-k\hbar\omega_\text{r}$, i.e.\ at the bias voltage where $\Gamma_\text{e}^+(E)$ is resonant. We assume that $k_BT \ll \epsilon$, such that the Fermi function can be approximated by a step function. In order to observe this peak, the following conditions have to be fulfilled:
\begin{enumerate}
    \item The tunneling rate $\Gamma_\text{e}^+(E)$ is resonant, i.e.\ $eV=-(\Delta-\epsilon)-k\hbar\omega_\text{r}$
    \item The other tunneling rate $\Gamma_\text{h}(E)$ is nonzero, i.e.\  $|\epsilon+eV+l\hbar\omega_\text{r}|>\Delta$.
    \item The difference in Fermi functions is nonzero, i.e.\ $|f(\epsilon-eV-k\hbar\omega_\text{r})-f(\epsilon+eV+l\hbar\omega_\text{r})|=1$.
\end{enumerate}
Combining the first condition with the other two conditions yields
\begin{equation}
    k>\frac{2\epsilon}{\hbar\omega_\text{r}}+l
\end{equation}
Owing to the second condition, we approximate $\Gamma_\text{h}(E)$ by a constant, i.e. $\langle \Gamma_\text{h} \rangle=\Gamma_\text{h}(E\gg \Delta)$. Applying these conditions to Eq.\ \eqref{eq:ardelta}, we find for the excited state electron tunneling current
\begin{equation}
    I_\text{ex,e}(V)=-\frac{e}{\hbar\eta_\text{R}} \sum_{l=-\infty}^{\infty}\sum_{k>\frac{2\epsilon}{\hbar\omega_\text{r}}+l} J^2_{k}\left(\alpha\right)J^2_{l}\left(\alpha\right)\Gamma_e^+(\epsilon-eV-k\hbar\omega_\text{r})\langle\Gamma_h \rangle 
    \label{eqn:excommae}
\end{equation}
In analogy to the Tien-Gordon model, we define a bare tunneling current which does not involve the modulation by the microwaves $I_\text{ex,e}^0(V)=- \frac{e}{\eta_s\hbar}\langle\Gamma_h\rangle\Gamma_e^+\left(\epsilon-eV\right)$. Equation \eqref{eqn:excommae} simplifies to
\begin{equation}
    I_\text{ex,e}(V)=\sum_k w(\alpha,k)J_k^2(\alpha)I_\text{ex,e}^0(V+k\hbar\omega_\text{r}/e),
\end{equation}
where we have defined the weight function $w(\alpha,k)$ as
\begin{equation}
    w(\alpha,k)=\sum_{m>m_0-k} J_m^2(\alpha)
\end{equation}
and where $m_0=\left \lceil{\frac{2\epsilon}{\hbar\omega}}\right \rceil$, where $\lceil\rceil$ is the ceiling function ($\lceil x \rceil$ is defined as $x$ rounded to the next larger integer). The weight function does not change the position nor the number of the replicas. It only modifies the amplitude of the peak. This nicely explains the appearance of replica at integer multiples of $\hbar\omega/e$ instead of $\hbar\omega/2e$. The weight function also introduces a threshold through the condition $m>m_0-k$, which means that $m_0$ quanta of $\hbar\omega$ have to be absorbed from the microwave in order to excite the YSR state. The leading edge of the weight function determining the onset of the peak as function of microwave intensity is given by the lowest order Bessel function $J^2_{m_0-k}(\alpha)$. This means in particular that the bare tunneling current $I_\text{ex,e}^0(V)$ as defined above cannot be observed when the microwave is turned off. 

\subsection{Simplified tunneling equations for ground state and excited state tunneling}

We can straightforwardly extend the above derivation for all four peaks. We find for the bare tunneling currents
\begin{eqnarray}
I_\text{ex,e}^0(V)&=&- \frac{e}{\eta_\text{R}\hbar}\Gamma_e^+\left(\epsilon-eV\right)\langle\Gamma_h\rangle,\\
I_\text{ex,h}^0(V)&=&+ \frac{e}{\eta_\text{R}\hbar}\langle\Gamma_e\rangle\Gamma_h^+\left(\epsilon+eV\right),\\
I_\text{gr,e}^0(V)&=&+ \frac{e}{\eta_\text{R}\hbar}\Gamma_e^-\left(\epsilon-eV\right)\langle\Gamma_h\rangle,\\
I_\text{gr,h}^0(V)&=& -\frac{e}{\eta_\text{R}\hbar}\langle\Gamma_e\rangle\Gamma_h^-\left(\epsilon+eV\right),
\end{eqnarray}
where the first index (gr,ex) refers to ground state and excited state tunneling and the second index (e,h) refers to electron and hole tunneling, respectively. From these bare tunneling currents, which have one peak each, we find the following equations to calculate the spectra with microwaves
\begin{eqnarray}
I_\text{ex,e}(V)&\approx&\sum_k w(\alpha,k)J_k^2(\alpha)I_\text{ex,e}^0(V+k\hbar\omega_\text{r}/e),\\
I_\text{ex,h}(V)&\approx&\sum_k w(\alpha,k)J_k^2(\alpha)I_\text{ex,h}^0(V
-k\hbar\omega_\text{r}/e),\\
I_\text{gr,e}(V)&\approx&\sum_k \tilde{w}(\alpha,k)J_k^2(\alpha)I_\text{gr,e}^0(V
-k\hbar\omega_\text{r}/e),\\
I_\text{gr,h}(V)&\approx&\sum_k \tilde{w}(\alpha,k)J_k^2(\alpha)I_\text{gr,h}^0(V
+k\hbar\omega_\text{r}/e),
\end{eqnarray}
where the weight functions are defined as
\begin{eqnarray}
w(\alpha,k)&=&\sum_{m\geq m_0-k}J_m^2(\alpha),\\
\tilde{w}(\alpha,k)&=&\sum_{m\geq-m_0-k}J_m^2(\alpha).
\end{eqnarray}
where $m_0=\left \lceil{\frac{2\epsilon}{\hbar\omega_\text{r}}}\right \rceil $ is the minimum number of quanta needed to excite the YSR state. Interestingly, we find that for ground state tunneling the weight function $\tilde{w}(\alpha,k)$ does not impose a threshold for the activation of the tunneling process, because the condition $m\geq-m_0$ (for $k=0$) always includes the zeroth order Bessel function, such that resonant Andreev processes are always possible without microwaves as has been discussed before \cite{si_Ruby2015}. 

\subsection{Quasiparticle tunneling from the ground state}

For completeness, we note that quasiparticle tunneling has to be considered, when modeling ground state tunneling, since the lifetime of the YSR state is not infinite in practice. In the deep tunneling regime, quasiparticle tunneling can be calculated from the Tien-Gordon model
\begin{equation}
    \centering
    I_\text{qp}\left(V,\alpha\right)=\sum_n  J_n^2\left(\alpha\right) I\left(V+ n\hbar\omega_\text{r}/e,0\right).
    \label{eqn:qp}
\end{equation}
As Andreev processes become more dominant with increasing tunneling conductance, the quasiparticle current reduces (cf.\ also regularized quasiparticle current in Eq.\ \eqref{eq-I_qp-reg}) \cite{si_Ruby2015,si_Huang2020}. Excited state tunneling is a two-electron tunneling process, so that quasiparticle tunneling does not apply for that process. The full Green's function model naturally includes all current contributions.

\section{Resonant vs.\ Nonresonant Andreev Processes}

In the previous section, we have derived a simple model that finds a spacing of $\hbar\omega/e$ between the replica of the resonant Andreev processes despite two charges being transferred. By contrast, the replica of nonresonant Andreev reflections are spaced by $\hbar\omega/2e$. We can explain this difference in behavior by deriving a simplified equation for the regular Andreev reflection starting from the same Eq.\ \eqref{eq-I_AR} as for the resonant Andreev processes. The main difference is that the anomalous Green's function $(g_\text{R})_{12}(E)$ is no longer given by a resonance, but by the standard result  
\begin{equation}
    (g_\text{R})_{12}(E) =\rho_\text{R}\frac{\Delta_\text{R}}{\sqrt{\Delta_\text{R}^2-\omega^2}},
\end{equation}
which in the following we will approximate by a constant, such that $|(g_\text{R})_{12}(E)|^2=\rho_\text{R}^2$. This is justified because the relevant part of the anomalous Green's function that is probed here is very close to zero energy. We further define
\begin{equation}
    \tilde{\Gamma}_\textrm{e}(E)  =  \tilde{\Gamma}_\textrm{h}(E) = 2 \pi |t|^2 \rho_\textrm{L}(E) \rho_\text{R},	
\end{equation}
Equation \eqref{eq-I_AR} changes for nonresonant Andreev reflections into
\begin{widetext}
\begin{equation} 
I_\textrm{AR}(V) = \frac{2e}{h} \sum_{k,l} J^2_k(\alpha) J^2_l(\alpha) \times 
\int^{\infty}_{-\infty} \tilde\Gamma_\textrm{e}(E-eV-k\hbar \omega_\text{r}) \, \tilde\Gamma_\textrm{h}(E+eV+l\hbar \omega_\text{r})\left[f(E-eV-k\hbar \omega_\text{r}) - f(E+eV+l\hbar \omega_\text{r}) \right] \, dE .
\label{eqn:nrar}
\end{equation}
\end{widetext}
For simplicity, we approximate the coherence peaks in $\rho_\text{L}(E)$ by very sharp Lorentzians at $E = \pm \Delta$ with a very small width $\eta_\text{L}$, which can be easily integrated. The Fermi functions do not impose any restrictions here. We find
\begin{equation} 
I_\textrm{AR}(V) = \frac{\pi e}{h\eta_\text{L}} \sum_{k,l} J^2_k(\alpha) J^2_l(\alpha)\frac{4\pi^2|t|^4\rho_\text{L}^2\rho_\text{R}^2}{(eV-\Delta+(k+l)\frac{\hbar\omega_\text{r}}{2})^2+\eta_\text{L}^2},
\label{eqn:nrarres}
\end{equation}
which is again a Lorentzian that nicely shows how the replica are spaced by $\hbar\omega/2e$. This demonstrates that depending on the presence of a resonance inside the superconducting gap, the spacing between replica changes accordingly. In this case, the spacing between replica cannot be used for inferring the number of charges being transferred.

\end{document}